\documentclass{aa}  
\usepackage{natbib}
\usepackage{comment}
\usepackage{graphicx}

\usepackage{txfonts}
\usepackage[dvipsnames]{xcolor}

\usepackage[breaklinks, colorlinks, citecolor=blue]{hyperref}

\newcommand{\A}{$A_{2}$}
\newcommand{\tb}{$t_{\rm bar}$}
\newcommand{\ls}{$l_{*}$}
\newcommand{\lgas}{$l_{g}$}
\newcommand{\Msun}{$\mbox{M}_{\sun}$}
\newcommand{\sigz}{$\sigma_{z}$}
\newcommand{\sigrel}{$\sigma_{\rm rel}$}
\newcommand{\vesc}{$v_{\rm esc}$}
\newcommand{\om}{$\Omega$}
\newcommand{\omb}{$\Omega_{b}$}
\newcommand{\kms}{km s$^{-1}$}

\begin{document}
    \title{Simulating nearby disc galaxies on the main star formation sequence}
    \subtitle{II. The gas structure transition in low  and high stellar mass discs}
  
   \author{
   Pierrick Verwilghen \inst{\ref{origins},\ref{eso}} \and
   Eric~Emsellem\inst{\ref{eso},\ref{lyon}} \and
   Florent Renaud\inst{\ref{sxb},\ref{usias}} \and
   Oscar Agertz\inst{\ref{Lund}} \and
   Milena Valentini\inst{\ref{UniTs},\ref{OATs},\ref{ICSC}} \and
   Amelia Fraser-McKelvie\inst{\ref{eso}} \and
   Sharon Meidt\inst{\ref{SOU}} \and
   Justus Neumann\inst{\ref{MPIA}} \and
   Eva Schinnerer\inst{\ref{MPIA}} \and
   Ralf S.\ Klessen\inst{\ref{ITA},\ref{IWR},\ref{CFA},\ref{Rad}} \and 
   Simon C. O. Glover\inst{\ref{ITA}} \and
   Ashley.~T.~Barnes\inst{\ref{eso}} \and
   Daniel~A.~Dale\inst{\ref{UWYO}} \and
   Damian R. Gleis\inst{\ref{MPIA}} \and
   Rowan J.\ Smith\inst{\ref{StA}} \and
   Sophia K. Stuber\inst{\ref{MPIA}} \and
   Thomas G. Williams\inst{\ref{Ox}}
   }
   
   \institute{
        Excellence Cluster ORIGINS, Boltzmannstraße 2, 85748 Garching, Germany\label{origins}
        \and European Southern Observatory, Karl-Schwarzschild-Stra{\ss}e 2, 85748 Garching, Germany\label{eso}
        \and Univ Lyon, Univ Lyon1, ENS de Lyon, CNRS, Centre de Recherche Astrophysique de Lyon UMR5574, F-69230 Saint-Genis-Laval France\label{lyon}
        \and Observatoire Astronomique de Strasbourg, Universit\'e de Strasbourg, CNRS UMR 7550, F-67000 Strasbourg, France\label{sxb}
        \and University of Strasbourg Institute for Advanced Study, 5 all\'ee du G\'en\'eral Rouvillois, F-67083 Strasbourg, France \label{usias}
        \and Lund Observatory, Division of Astrophysics, Department of Physics, Lund University, Box 43, SE-221 00 Lund, Sweden\label{Lund}
        \and Astronomy Unit, Department of Physics, University of Trieste, via Tiepolo 11, I-34131 Trieste, Italy\label{UniTs}
        \and INAF – Osservatorio Astronomico di Trieste, via Tiepolo 11, I-34131 Trieste, Italy\label{OATs}
        \and ICSC - Italian Research Center on High Performance Computing, Big Data and Quantum Computing\label{ICSC}
        \and Sterrenkundig Observatorium, Universiteit Gent, Krijgslaan 281 S9, B-9000 Gent, Belgium\label{SOU}
        \and Max Planck Institut f\"ur Astronomie, K\"onigstuhl 17, 69117 Heidelberg, Germany\label{MPIA}
        \and Universit\"{a}t Heidelberg, Zentrum f\"{u}r Astronomie, Institut f\"{u}r Theoretische Astrophysik, Albert-Ueberle-Str. 2,\\ D-69120 Heidelberg, Germany \label{ITA} 
        \and Universit\"{a}t Heidelberg, Interdisziplin\"{a}res Zentrum f\"{u}r Wissenschaftliches Rechnen, Im Neuenheimer Feld 205,\\ D-69120 Heidelberg, Germany\label{IWR} 
        \and Center for Astrophysics, Harvard \& Smithsonian, 60 Garden Street, Cambridge MA, USA\label{CFA}
        \and Elizabeth S. and Richard M. Cashin Fellow at the Radcliffe Institute for Advanced Studies at Harvard University, 10 Garden Street, Cambridge, MA 02138, USA \label{Rad}
        \and Department of Physics and Astronomy, University of Wyoming, Laramie, WY 82071, USA\label{UWYO}
        \and SUPA, School of Physics and Astronomy, University of St Andrews, North Haugh, St Andrews, KY16 9SS, UK\label{StA}
        \and Sub-department of Astrophysics, Department of Physics, University of Oxford, Keble Road, Oxford OX1 3RH, UK\label{Ox}
   }

   \date{Received February, 2025; accepted May, 2025}

  \abstract
  {
   Recent  hydrodynamical simulations of isolated barred disc galaxies have suggested a structural change in the distribution of the interstellar medium (ISM) around a stellar mass M$_{*}$ of $10^{10}$~\Msun.  In the higher-mass regime (M$_{*} \geq 10^{10}$ \Msun), we observe the formation of a central gas and stellar disc with a typical size of a few hundred parsecs connected through lanes to the ends of the stellar bar. In the lower-mass regime (M$_{*} < 10^{10}$ \Msun), such an inner disc is absent and the gas component exhibits a more chaotic distribution. Observations of nearby star-forming galaxies support the existence of such a change. These inner gas discs may represent an important intermediate scale connecting the large kiloparsec-scale structures with the nuclear (sub-parsec) region, transporting gas  inwards to fuel the central supermassive black hole (SMBH). For this work we used an extended set of high-resolution hydrodynamical simulations of isolated disc galaxies with initial properties (i.e. stellar mass, gas fraction, stellar disc scale length, and the bulge mass fraction) with properties covering the range of galaxies in the PHANGS sample to investigate this change of regime. We studied the physical properties of the star-forming ISM in both stellar mass regimes and extracted a few physical tracers: the inner Lindblad resonance (ILR), the probability distribution function (PDF), the virial parameter, and the Mach number. In line with observations, we confirm a structure transition in the simulations that occurs between a stellar mass of $10^{9.5}$ and $10^{10}$ \Msun. We show that the physical origin of this change of regime is driven by stellar feedback and its contribution relative to the underlying gravitational potential. With their shallower potential and typically higher gas mass fraction, lower-mass disc PHANGS galaxies combine two ingredients that significantly delay or even prevent the formation of a central gas (and stellar) disc. These results shed some light on the observed properties of star-forming barred galaxies and have implications for the star formation regimes, the growth of central structures, and the overall secular evolution of disc galaxies.}
   

   \maketitle

\nolinenumbers
\section{Introduction}\label{sec:intro}
Recent observational campaigns \citep[see e.g. PHANGS\footnote{PHANGS: Physics at High Angular resolution in Nearby GalaxieS (https://sites.google.com/view/phangs/home)};][]{Leroy2021, Lee+2022, Emsellem2022, Lee+2023} have emphasised  the fact that nearby disc galaxies exhibit a wide variety of structures in the distribution of the interstellar medium (ISM) such as spiral arms, dust lanes, and inner gas discs \citep{Querejeta2021,Stuber+2023}. One of the most common characteristics is the presence of a well-defined gas concentration \citep[rings or discs; see e.g.][]{Knapen2005,Comeron2010,Stuber+2023,Erwin2024} within the central few hundred parsecs of galactic bars, often referred to as central molecular zones (CMZs), to echo the observed flattened structure in the Milky Way \citep[see][]{Morris96,Henshaw2023}. These disc-like structures are sometimes believed to represent an intermediate stalling region towards a potential supermassive black hole (SMBH), and can thus be considered as `gas reservoirs', a small fraction of which could help feed the central SMBH \citep{Emsellem2015, Robichaud+17, Yu2022}.

The characteristic arrangement of the stellar bar, dust and gas lanes, and central disc reservoir seems to be preferentially observed in relatively high stellar mass barred disc galaxies \citep[see e.g.][]{DiazGarcia2020}. A stellar bar is an essential ingredient that drives the inward gas fuelling from kiloparsec scales to the inner region, often via higher-density gas lanes that connect with the inner gas reservoirs. These reservoirs are naturally associated with an increase in the local surface density of the ISM and may thus catalyse the formation of new stars \citep{Sormani2020, Schinnerer2023}. Star formation (SF) further leads to stellar-driven feedback, one of the mechanisms that may also influence the gas transport  down to sub-parsec scales.

\begin{figure}[ht]
\centering
\includegraphics[width=0.5\textwidth]{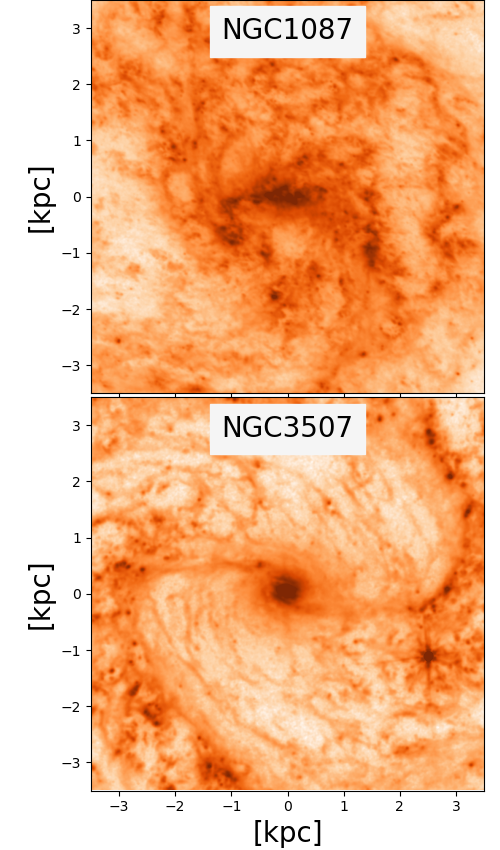}
\caption{JWST MIRI 7.7~$\mu$m images of two nearby barred main-sequence star-forming galaxies (GO 3707; PI Leroy) NGC\,1087 and NGC\,3507 (stellar masses of $\sim 10^{9.95}$ and $\sim 10^{10.4}$~\Msun, respectively) emphasising the difference in morphology. Both images have been deprojected, and the bar is set horizontally. The field of view is 7~kpc\,$\times$\,7~kpc, thus showing the central $\pm 3.5$~kpc. These data have previously appeared in \citet{Chown2024} with data reduction following \citet{Williams2024}.}
\label{fig:2gal}
\end{figure}

Observations suggest that these gas reservoirs only emerge in barred galaxies with a stellar mass above a given threshold \citep[e.g.][]{Verley2007, FraserMcKelvie2020}. Most nearby disc galaxies on the star formation main sequence (SFMS) above a stellar mass of $\sim 10^{10}$~\Msun\ have gas mass fractions below 25\% \citep{Catinella2012,Butcher2016} and the neutral, ionised, and molecular ISM is well structured around a well-defined galactic centre (see bottom panel of Fig.~\ref{fig:2gal}). In systems with stellar mass below $10^{10}$~\Msun, the gas inside the bar region often appears less structured, and its surface density distribution is more complex and less regular (see top panel of Fig.~\ref{fig:2gal}). The gas mass fraction is also significantly higher than in more massive discs, reaching values above 40\%, and the structure of their ISM appears more clumpy and disorganised, with star-forming regions located all along the stellar bar. Observations suggest that the galaxy's stellar mass is a key parameter in determining whether a disc galaxy forms a central gas reservoir. While the properties of gas reservoirs in higher-mass systems have been investigated via observations \citep{Sormani2019, Neumann2019, Schinnerer2023, Choi2023, Sormani2023B} and simulations \citep{Atha1992gas, Seo2019, Sormani2024}, the large-scale galactic or small-scale local physical conditions required for the emergence of a central gas disc or ring are still not well understood.

For this work we investigated the time evolution of main-sequence star-forming discs in order to better understand the underlying principles that drive the formation and growth of central structures. We addressed  whether or not this evolution depends on stellar mass, for example, and questioned the origin of this dependence, using controlled experiments, i.e. simulations mocking star-forming main-sequence galaxies. We thus built an extended grid of generic disc galaxies \citep{Verwilghen_2024} based on properties of nearby disc galaxies on the SFMS \citep{Leroy2019}. We used various physical tracers, for example the inner Lindblad resonance (ILR), the probability density function (PDF), the virial parameter, and the Mach number to probe a potential change of regime in the SF distribution, focusing on the advent of bar-driven central mass concentrations.

After briefly presenting the extended set of hydrodynamical simulations, we illustrate in Section~\ref{sec:illustration} the overall variation in the evolutionary tracks and properties of galaxies in the low- and high-mass regimes. We focus on the properties of the gas before and after the formation of the bar in Section~\ref{sec:tracers}, using tracers such as the velocity dispersion, the Mach number, or the virial parameter. In Section~\ref{sec:feedback_grav} we further demonstrate the role of stellar feedback in establishing such a change of regime, and its relative weight against the local gravitational potential. Section~\ref{sec:conclusion} follows with a brief discussion and a summary of the results.

\section{First evidence in simulations for a change of regime}
\label{sec:illustration}
In this section, we emphasise the differences in evolution of the gas morphology and distribution of star-forming regions (Sect.~\ref{sec:gasdistrib}) between the low  ($<10^{10}$ \Msun) and high  ($\geq10^{10}$ \Msun) stellar mass models within our set of simulations (briefly presented in Sect.~\ref{sec:set_simus}). We also illustrate the variation of the surface density profiles below and above this stellar mass threshold.
\subsection{The set of simulations}
\label{sec:set_simus}
In the first paper of this series \citep{Verwilghen_2024}, we used our first set of 16 isolated simulated galaxies to analyse the evolution of the gas mass inside the central 1~kpc region and its connection with the bar. We identified four phases in the bar-driven fuelling of gas towards the central 1~kpc region. These four phases are accompanied by a central starburst and are only observed in the higher stellar mass models (i.e. M$_{\star}$ $\leq 10^{10}$ \Msun), which are all located on the SFMS. This starburst is never observed in the lower-stellar mass models (i.e. M$_{\star}$ = $10^{9.5}$ \Msun). Additionally, we found that gas reservoirs are only formed in higher stellar mass discs, showing an organised gas structure with star-forming regions mainly concentrated inside the central reservoir. In lower-stellar mass discs, the gas structure is turbulent, and the star-forming regions are distributed along the bar.

In this paper, we further probe these processes using an extended set of high-resolution simulations of isolated nearby disc galaxies based on the properties of the PHANGS galaxy sample \citep{Verwilghen_2024}. This simulation set is currently composed of a grid of 35 models having different control parameters encoded as GxxxMxxxFxxLxBxx, where the number after G stands for an internal numbering reference, M for the stellar mass (i.e. 095, 100, 105, and 110 for the stellar mass bins $10^{9.5}$, $10^{10}$, $10^{10.5}$, and $10^{11}$ \Msun, respectively), F for the total gas fraction (i.e. 10, 20, and 40\%), L for the scale length of the stellar disc (2-5 kpc), and B for the bulge mass fraction (i.e. 0, 10 and 30 \% of the total stellar mass). As mentioned above, a first sub-set of 16 simulations with a more restricted set of parameters has been presented in \cite{Verwilghen_2024}, with an emphasis on the phases and timescales associated with bar formation. In Table~\ref{Tab:Grid_35}, we list the values of the control parameters of the initial conditions for all 35 models in the present extended set.

Briefly, all simulations have been performed with RAMSES \citep{Teyssier2002}, an adaptive mesh refinement (AMR) hydrodynamical code. The code treats the stars and dark matter (DM) as particles and gas with cells via the AMR grid. The code also includes numerical recipes implementing the cooling of gas, SF and stellar feedback (supernovae and stellar winds), and the evolution of the metallicity via two tracers \citep[see][for more details]{Agertz2013,Agertz2021}. Those simulations reach a minimum spatial sampling of 12~pc, and a typical mass resolution of $10^4$ and $2\times 10^3$~\Msun\ for the old and new stellar particles, respectively. More details regarding the sub-grid recipes, detailed setup of the simulations and initial conditions can be found in \cite{Verwilghen_2024}. 
The radial size of our simulated bars within the first 3~Gyr typically ranges between 1 and 1.5 \ls\ (the scale length of the main stellar disc), thus roughly scaling with the stellar mass of the host.
\begin{table}[ht]
\caption{List of all models in the extended simulation set.}
\centering
\resizebox{\columnwidth}{0.85\columnwidth}{%
\begin{tabular}{|c|c|c|c|c|c|c|} 
\hline
Model & $\log_{10}(M_{\star}$) & F & \ls & $l_g$/\ls & B & BAR \\
 & [M$_{\odot}$] & [\%] & [kpc] & & [\%] &  \\
\hline
G001M095F10L2B00 & 9.5 & 10 & 2 & 2 & 0 & Y \\
\hline
G002M095F10L2B10 & 9.5 & 10 & 2 & 2 & 10 & Y \\
\hline
G007M095F10L3B00 & 9.5 & 10 & 3 & 2 & 0 & N \\
\hline
G008M095F10L3B10 & 9.5 & 10 & 3 & 2 & 10 & N \\
\hline
G009M095F10L3B30 & 9.5 & 10 & 3 & 2 & 30 & N \\
\hline
G013M095F20L2B00 & 9.5 & 20 & 2 & 2 & 0 & Y\\
\hline
G014M095F20L2B10 & 9.5 & 20 & 2 & 2 & 10 & Y\\
\hline
G015M095F20L2B30 & 9.5 & 20 & 2 & 2 & 30 & N \\
\hline
G019M095F20L3B00 & 9.5 & 20 & 3 & 2 & 0 & N \\
\hline
G020M095F20L3B10 & 9.5 & 20 & 3 & 2 & 10 & N \\
\hline
G021M095F20L3B30 & 9.5 & 20 & 3 & 2 & 30 & N \\
\hline
G025M095F40L2B00 & 9.5 & 40 & 2 & 2 & 0 & Y\\
\hline
G026M095F40L2B10 & 9.5 & 40 & 2 & 2 & 10 & Y\\
\hline
G027M095F40L2B30 & 9.5 & 40 & 2 & 2 & 30 & Y\\
\hline
G031M095F40L3B00 & 9.5 & 40 & 3 & 2 & 0 & Y\\
\hline
G032M095F40L3B10 & 9.5 & 40 & 3 & 2 & 10 & N \\
\hline
G033M095F40L3B30 & 9.5 & 40 & 3 & 2 & 30 &  N\\
\hline
G037M100F10L2B00 & 10 & 10 & 2 & 2 & 0 & Y\\
\hline
G038M100F10L2B10 & 10 & 10 & 2 & 2 & 10 & Y\\
\hline
G039M100F10L2B30 & 10 & 10 & 2 & 2 & 30 & Y\\
\hline
G045M100F10L3B00 & 10 & 10 & 3 & 2 & 0 & N \\
\hline
G046M100F10L3B10 & 10 & 10 & 3 & 2 & 10 & N \\
\hline
G053M100F20L2B00 & 10 & 20 & 2 & 2 & 0 & Y\\
\hline
G054M100F20L2B10 & 10 & 20 & 2 & 2 & 10 & Y\\
\hline
G055M100F20L3B30 & 10 & 20 & 2 & 2 & 30 & Y\\
\hline
G069M100F40L2B00 & 10 & 40 & 2 & 2 & 0 & N \\
\hline
G070M100F40L2B10 & 10 & 40 & 2 & 2 & 10 & Y\\
\hline
G105M105F10L3B00 & 10.5 & 10 & 3 & 2 & 0 & Y\\
\hline
G106M105F10L3B10 & 10.5 & 10 & 3 & 2 & 10 & Y\\
\hline
G137M105F20L3B00 & 10.5 & 20 & 3 & 2 & 0 & Y\\
\hline
G138M105F20L3B10 & 10.5 & 20 & 3 & 2 & 10 & Y\\
\hline
G161M110F10L5B00 & 11 & 10 & 5 & 2 & 0 & Y\\
\hline
G162M110F10L5B10 & 11 & 10 & 5 & 2 & 10 &  N \\
\hline
G177M110F20L5B00 & 11 & 20 & 5 & 2 & 0 & Y\\
\hline
G178M110F20L5B10 & 11 & 20 & 5 & 2 & 10 & Y\\
\hline
\end{tabular}%
}
\tablefoot{Columns from left to right are: label encapsulating the parameters used to set up the initial conditions, total stellar mass ($M_{\star}$, here in $\log_{10}$), gas mass fraction F (in \%), stellar scale length \ls\ (in kpc), scale length ratio $l_g$/\ls\ (gas over stellar scale lengths), and bulge mass fraction B (in \%). The last column shows models that developed a bar (Y) or not (N) by the end of the run.}
\label{Tab:Grid_35}
\end{table}
\subsection{Gas surface density distribution}
\label{sec:gasdistrib}
\begin{figure*}[h!]
\centering
\includegraphics[width=\textwidth]{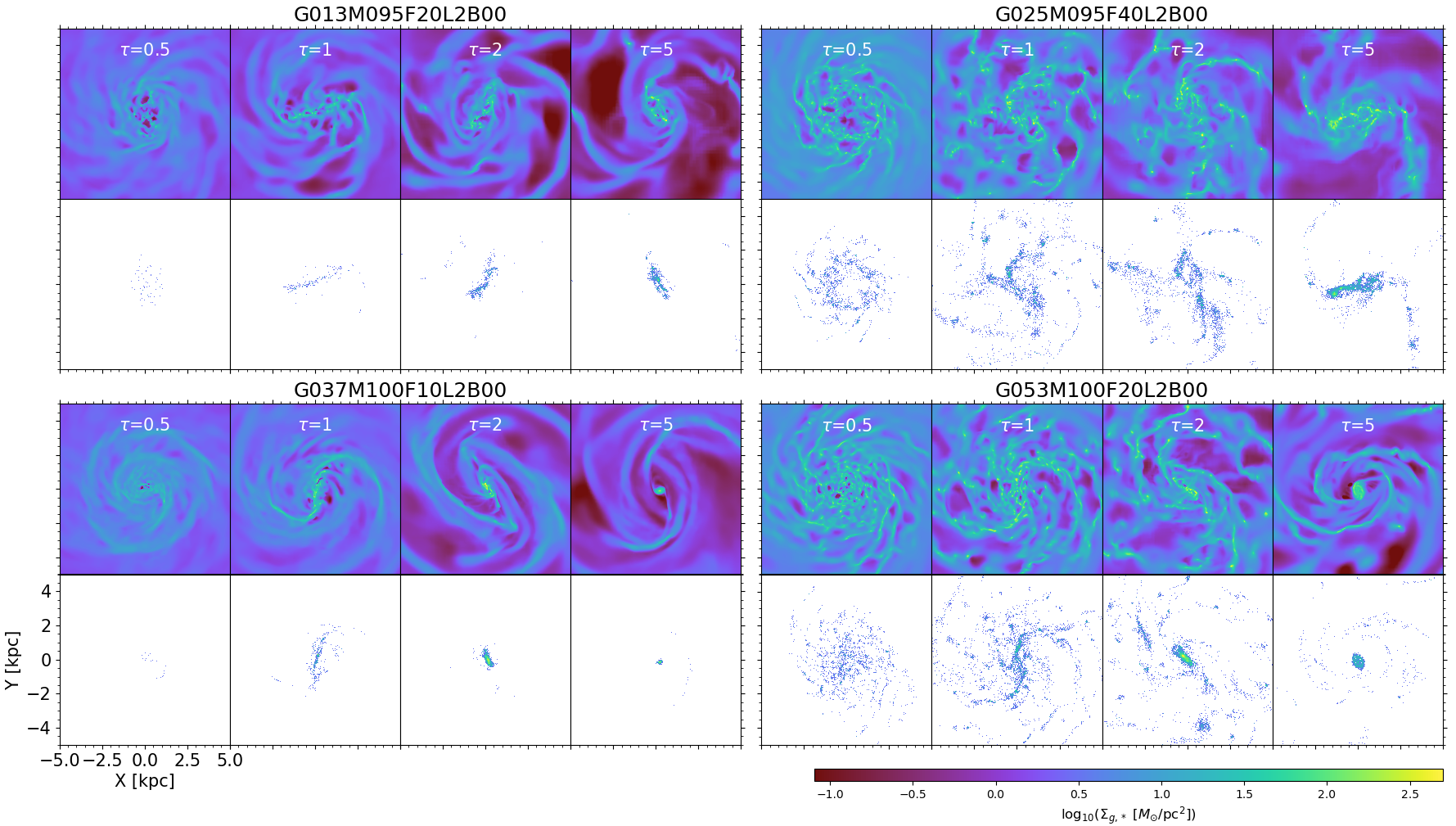}
\caption{Surface density map of gas and newly formed stars ($\leq 100$ Myr) of four models in two stellar mass bins (G013, G025, with an initial stellar mass of $10^{9.5}$~\Msun; G037, and G053 with an initial stellar mass of $10^{10}$~\Msun) from our set of simulations. Each panel shows one model at different values of the parameter $\tau$, with $\tau=1$ corresponding to the bar formation timescale. G013 and G053 have the same 20\% gas mass fraction, while G025 and G037 have a 40\% and 10\% gas mass fraction, respectively (the latter two following the averaged properties of star-forming main-sequence galaxies at their respective stellar masses).}
\label{fig:sig_gas_tau05-5}
\end{figure*}
A systematic visual inspection of all simulations listed in Table~\ref{Tab:Grid_35}
provides the first direct evidence for a change of regime: it is illustrated in Fig.~\ref{fig:sig_gas_tau05-5} in the central gas structures observed before and after the bar forms. In that figure, we show the evolution of the face-on gas surface density as a function of the dimensionless parameter $\tau$ (i.e. absolute $time$ normalised by the bar formation time \tb) for four models, in two stellar-mass bins (i.e. $10^{9.5}$ and $10^{10}$~\Msun -- top and bottom rows, respectively -- with two different gas fractions -- left and right panels). $\tau=1$ corresponds to the bar formation time set when $A_2$ reaches a threshold value of 0.2, with $A_2$ being the amplitude of the second Fourier term derived from the stellar density map \citep[see][]{Efstatiou1982, Atha_2_2002, Atha_2_2013}. The $A_2$ values we typically obtain for barred models range between 0.25 and 0.3, with their values increasing with the stellar mass. The typical bar formation time is of the order of $\sim400$ Myr and $800$ Myr for models without and with an additional central ellipsoid (i.e. the bulge), respectively \citep[see][for more details]{Verwilghen_2024}. Before the bar forms (i.e. $\tau = 0.5$), the gas distribution appears very clumpy and relatively unstructured, without any coherent large-scale gas structure. Once the stellar bar has formed, the gas distribution changes significantly, with the higher-density regions mostly clustered within the bar close to its major axis and along spiral-like structures outside the bar. 

Around $\tau = 2$, the evolution starts to differ between the low and high stellar mass models. The higher stellar mass models G037 and G053 ($10^{10}$~\Msun, with a gas fraction $\alpha$ of 10 and 20\%, respectively) exhibit well-organised gas surface density distributions. We observe the emergence of a central gas concentration in both G037 and G053, leading to the formation of a well-defined central vertically flattened overdensity at $\tau=2$. At $\tau = 5$, that structure has grown into an inner gas disc connected with both ends of the bar via distinct lanes.
We note that the size of the gas reservoir at $\tau=5$ is larger in G053 than in G037. Such a difference can be traced back to the earlier formation of the ILR in G053 (see Sect.~\ref{subsec:ilr}), following its initial higher gas fraction (20\%) and subsequent higher SF rate. The presence of such central gas mass concentrations significantly contrasts with the structures observed in the lowest-stellar mass ($10^{9.5}$ \Msun) models G013 and G025 ($\alpha$ of 20 and 40\%, respectively): the central gas structure remains very clumpy even after the bar forms \citep[see also][]{Bland2024}, and there is no apparent significant change in the gas distribution with time until $\tau = 5$ and beyond. 

At first glance, such a difference could be attributed to the higher gas fraction of models G013 and G025, which reflects the overall relative increased importance of gas in lower mass galaxies \citep[see, e.g.][]{Verwilghen_2024}. However, G013 and G053 are directly comparable: they are based on the same initial mass models for the stars, gas and dark matter, share the same gas fraction (20\%), G013's total mass only being scaled down by a factor of about 3 (or 0.5~dex). At $\tau = 5$, the lower mass model G013 still exhibits a complex gas distribution made of rapidly evolving clumps and voids within the bar region with no well-defined centre, while the higher-mass model G053 has a well-organised, nearly symmetric set of gas lanes fueling a central inner disc (the gas reservoir).
\begin{figure*}[h!]
\centering
\includegraphics[width=1\textwidth]{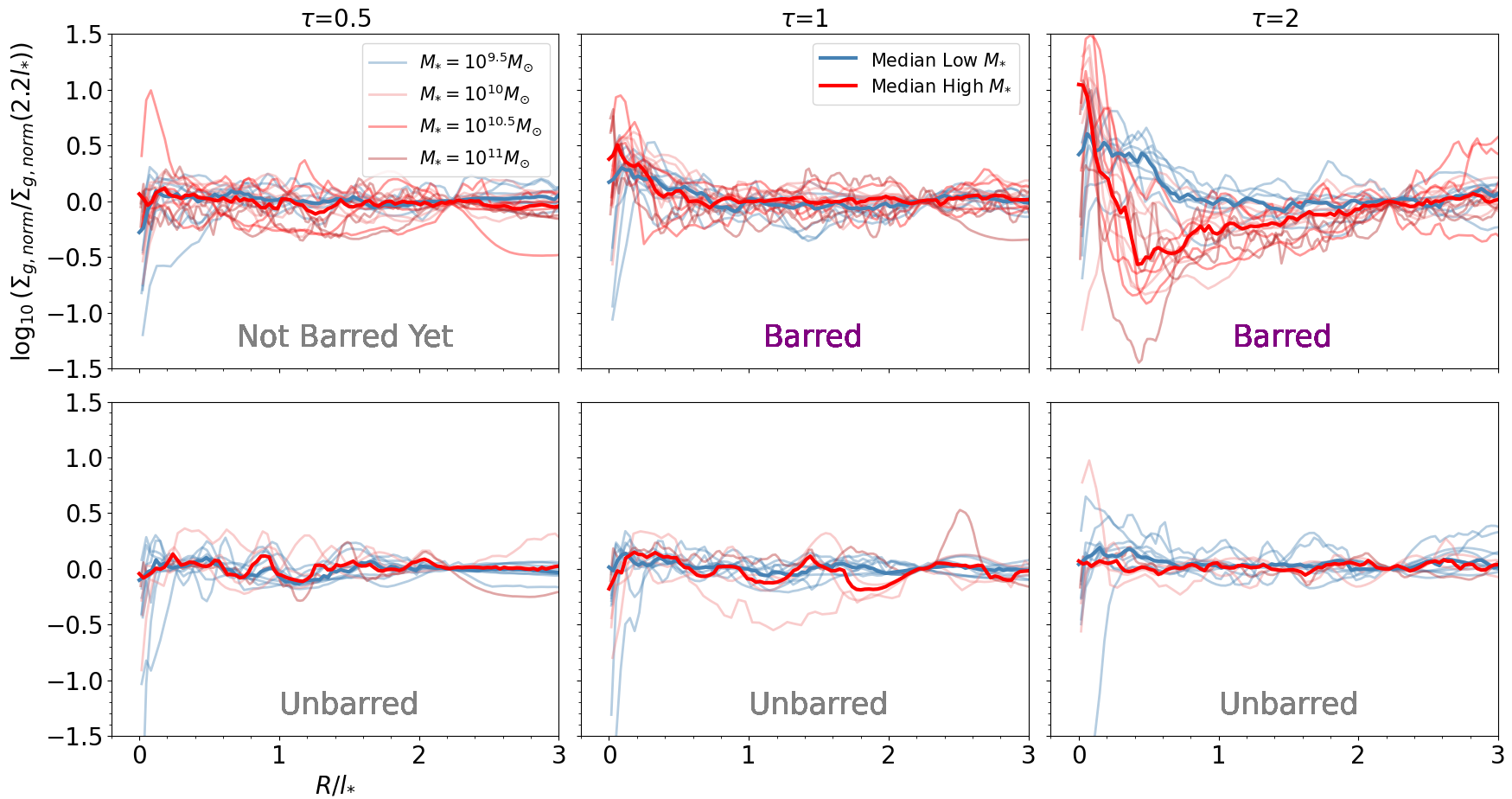}
\caption{Normalised surface gas density profiles $\Sigma_{g,norm}(\tau, R)$ for our set of simulations at $\tau$ = 0.5, 1, and 2 (see text; $\Sigma_{g,norm}(\tau, R) = \Sigma_g / \Sigma_g(\tau = 0, R)$). All $\Sigma_{g,norm}$ profiles have been divided by the value of $\Sigma_{g,norm}$ at $R = 2.2\cdot$\ls, for legibility. Each thin blue curve represents a low stellar mass model ($<10^{10}$ \Msun) and each thin red curve corresponds to a higher stellar mass model ($\geq 10^{10}$ \Msun; see inset  in the top left panel). The thick blue (resp. red) line in each panel is the associated median of the thin blue (resp. red) curve. The top panels show models that have formed a bar, sustained until the end of the simulation. The bottom panels illustrate the unbarred models (which are mostly those with large initial spheroids, e.g. G015 and G032, or extended initial discs, e.g. G043; see Appendix~\ref{sec:appa}).}
\label{fig:surf_tau05-5}
\end{figure*}
To provide a comparison of the evolution track between barred and unbarred models, we also provide in Fig~\ref{fig:gas_map_nobar} the gas and new star density maps for four unbarred models (G015, G032, G045, and G069) at four different time steps: 500, 1000, 1500, and 2000 Myr.  We see that simulated galaxies that do not host a bar form very few stars overall. We naturally do not observe the formation of a central gas reservoir in those unbarred models. The case of model G069 is special because a bar forms around 315 Myr, but is later destroyed by an incoming massive stellar cluster as shown in Fig.~\ref{fig:sf_map_G069} and is therefore considered as an unbarred model. In the rest of this work, the time corresponding to $\tau=1$ is artificially assigned to unbarred models using the \tb\ value of the barred model with similar initial conditions (e.g. model G007 is assigned the \tb\ value from G001, while G008 and G009 are assigned the \tb\ value from G002).
\subsection{The radial gas density re-distribution}
\label{sec:radialgas}
The presence or absence of an inner gas disc can be quantitatively illustrated by computing the radial gas density profiles from our simulations. Figure~\ref{fig:surf_tau05-5} shows the time evolution of the (azimuthal averaged) radial gas surface density profiles of the full set of models at $\tau$ = 0.5, 1, and 2. To emphasise any potential time changes, we have normalised all density profiles by the initial radial gas profile (at the start of the simulations, $t=0$). The profiles are plotted as a function of the galactic radius (R/\ls, where \ls\ is the initial stellar disc scale length). To ease the inter-model comparison, all profiles are finally rescaled by their respective values at the reference scaling of 2.2~\ls\ \citep[see e.g.][]{Courteau+2014}.\footnote{In our models, the typical scale length of the gas disc (\lgas) is twice the typical scale length of the stellar disc (\ls).} Figure~\ref{fig:surf_tau05-5} is split between barred galaxies (top panels) and unbarred ones (lower panels) using the classification at the end of the simulation ($t > 3$~Gyr). 

The profiles of most unbarred galaxies consistently keep a rather flat profile (within $\pm 0.2$~dex) at all times: the radial gas distribution does not seem to vary significantly over many 100s of Myr in the absence of a bar. Only one model departs significantly from this trend: G069. The bar in G069 forms around 315~Myr, but is subsequently ($t \sim 1500$~Myr) dissolved after interacting with an in-spiralling massive star cluster that has formed in the disc. Hence, this system is classified as unbarred by the end of the simulation, and its gas density profile becomes flat again after the bar has vanished, similar to other unbarred simulations.

This picture is radically different for barred galaxies as illustrated in the top panels of Fig.~\ref{fig:surf_tau05-5}. For barred galaxies, we first witness a central increase of the density profiles for all models by the time the bar has formed ($\tau = 1$).
At $\tau$=2, the lower- and higher-mass models exhibit a significantly different radial profile within the central $\sim 2$~\ls. The higher-mass galaxies show a strong relative drop with a minimum at $\sim 0.4$~\ls\ by a factor between 3 and 15, followed by a steep increase towards the centre (a factor of three of the initial, $t=0$, density). This contrasts with the lower-mass systems that show a slow steady rise from $\sim 2$~\ls\ inwards. We note that the maximum (normalised) central density values are also larger by about 0.5~dex on average for the more massive systems. The striking difference in evolution between the barred and unbarred systems emphasises the bar-driven re-organisation of the gas distribution. In systems with masses about $10^{10}$~\Msun\, this redistribution acts specifically, with the observed compact peaks within the central $\sim 0.2$~\ls\ being a witness of building a well-defined inner gas disc or `reservoir'. 

\subsection{The redistribution of SF across the disc}
The change observed in the evolution of the gas distribution mentioned above can be further associated with a redistribution of star-forming regions across the disc. In Fig.~\ref{fig:sig_gas_tau05-5} we illustrate this via a few snapshots of the new stars (with ages less than 100~Myr) formed during the simulation for the same four models (i.e. low stellar mass models G013 and G025, and high stellar mass models G037 and G053) and four values of $\tau$ 
(from 0.5 to 5). Low stellar mass models (top panels) have their star-forming regions mainly located along the stellar bar for model G013 and spread over the disc for model G025, with no emergence of a well-defined central nuclear disc structure. The evolution of the high stellar mass models (bottom panels) exhibits a clear transition in the distribution of star-forming regions, which is also observed in the highest-stellar mass bin models (i.e. $10^{10.5}$ and $10^{11}$ \Msun) developing a bar. At $\tau=2$, the star-forming regions are located along the bar as in the low stellar mass models. From $\tau=2$ to $\tau=5$, star-forming regions are mostly restricted to a central region with a typical size of a few hundred parsecs, associated with the formed central gas reservoirs (see the corresponding top row for each model).
\section{Physical tracers}
\label{sec:tracers}
In the previous section, we emphasised the formation of a central gas reservoir and the associated central distribution of star-forming regions in high stellar mass models ($\geq 10^{10}$ \Msun), and its absence in the lower-mass systems, by visually inspecting the gas and stellar surface densities. In this section, we turn to a few physical properties of the gas to quantitatively follow the emergence of the central mass concentration, then examine the gas density probability distribution function (PDF), the virial parameter and the Mach number inside the central 1~kpc region as tracers of the physical state of the gas.
\subsection{Emergence of the ILR} \label{subsec:ilr}
\begin{figure*}[h!]
    \centering
    \includegraphics[width=1\textwidth]{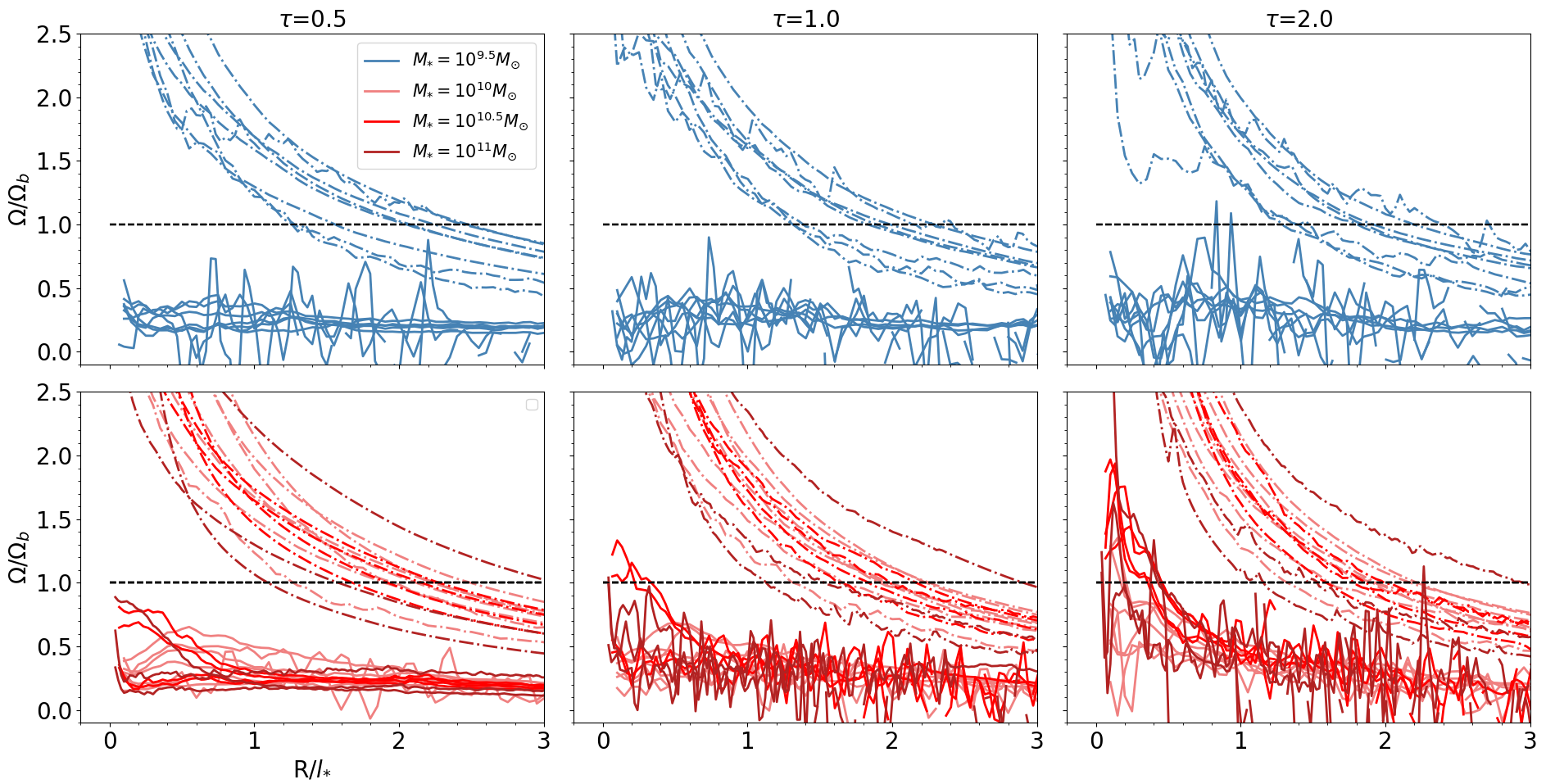}
    \caption{Evolution over time (from left to right, $\tau = 0.5, 1$, and 2) of the normalised angular frequency $\Omega$ (dot-dashed lines) and $\Omega-\kappa/2$ (solid curves) radial profiles of the barred low stellar mass (blue curves, top row) and high stellar mass (red curves, bottom row) models. The quantity   used in the normalisation is the bar pattern speed $\Omega_{b}$. The horizontal black dashed line represents $\Omega=\Omega_{b}$.}
    \label{fig:ILR}
\end{figure*}

In Sect.~\ref{sec:radialgas}, we showed that a change of regime is associated with the build-up of a central gas mass reservoir, accompanied by a corresponding central stellar overdensity made of newly formed stars. Such a change in the radial mass profile should be reflected in a steepening of the circular velocity and angular frequency (\om) profiles. A bar rotating with a given pattern speed (\omb) implies the potential emergence of inner resonances, one of the most commonly studied being the (-2, 1) ILR when $\Omega_b = \Omega - \kappa / 2$, with $\kappa$ the radial epicyclic frequency. In Fig.~\ref{fig:ILR}, we thus probe for all the barred models the time evolution of the angular frequency \om\ (top dashed-dotted curves) and the quantity $(\Omega-\kappa/2)$ (bottom curves) normalised by the pattern speed of the bar \omb.

At $\tau=0.5$, before the bar has formed, nearly all $\Omega - \kappa/2$ profiles are flat and all stay below the $\Omega_b$ value ($\Omega / \Omega_b = 1$; dashed lines). A few models exhibit a slight increase towards the centre, and those are the systems with an initially added ellipsoid (flattened bulge): this is consistent with the bulge representing an extra mass concentration, but has no relation with a bar-driven component.
At $\tau=1$, a few high-mass systems exhibit central increases that cross the \om/\omb\,$= 1$ line: an ILR is present. At late times ($\tau=2$), most of the higher-mass systems, including those that did not have an initial bulge, have a well-defined ILR. This result is consistent with the presence of an inner gas and stellar disc in the central region of those models. The models to first cross this threshold have a bulge in their initial conditions, hence starting with a steeper central potential: they form a central gas reservoir at earlier $\tau$ values. This dramatically contrasts with all of the lower-mass models that are far from having the inner mass concentration that would lead to an ILR: the lower-mass systems were not able to form a central disc-like reservoir of gas (or stars). The physical sizes of the ILR observed in our barred models at $\tau = 5$ range from $\sim 0.6$ to 1.5~kpc, within the interval of estimated values in the PHANGS sample \citep[see][]{Ruiz-Garcia2024}. Those values are also at the upper end of the sizes of inner rings \citep{Erwin2024}, as expected from theoretical considerations \citep{Sormani2023}.
\subsection{The central PDF}
\label{sec:pdf}
\begin{figure}[h!]
\centering
\includegraphics[width=0.5\textwidth]{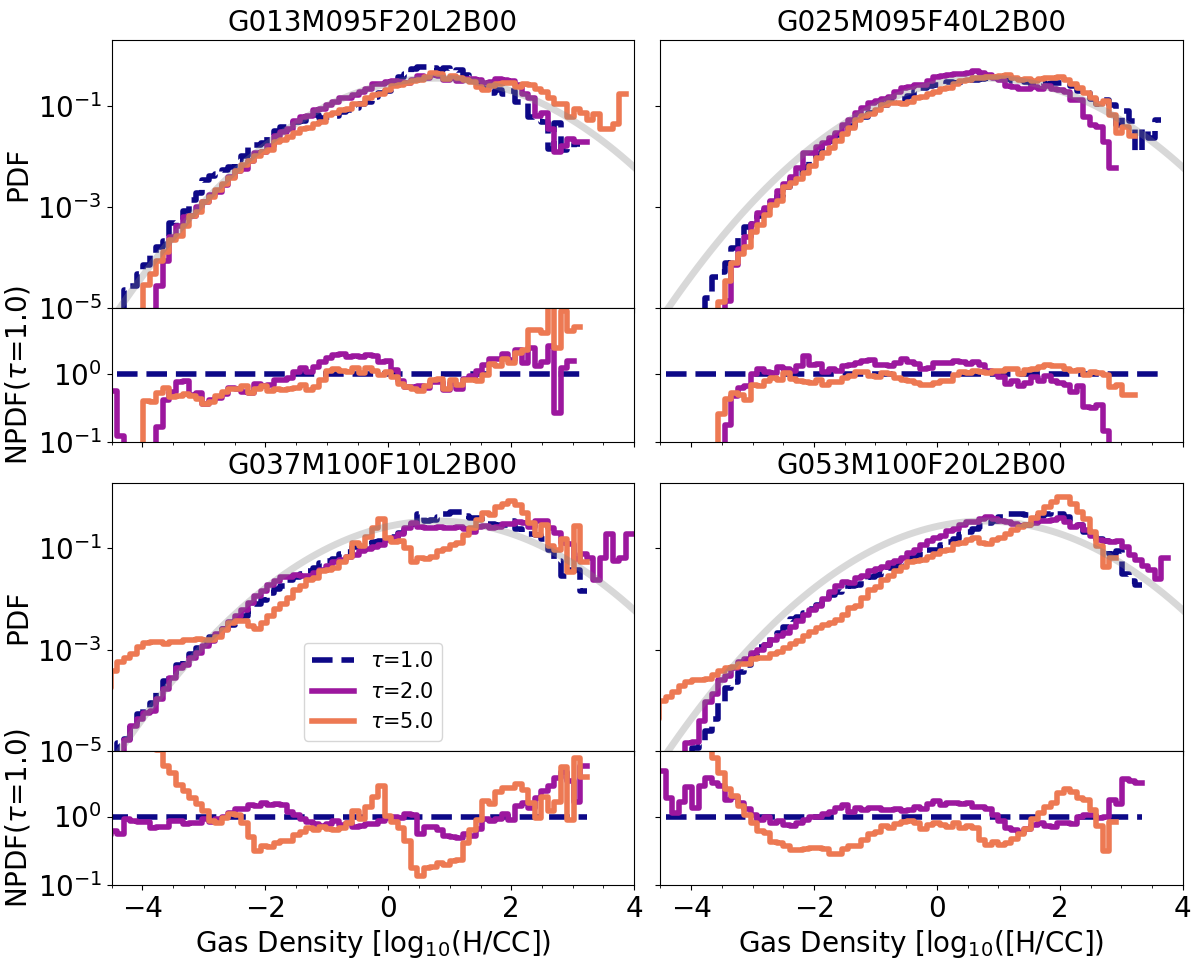}
\caption{Evolution of the mass-weighted gas density PDFs as a function of $\tau$ (for $\tau=1$, 2, and 5) of four of our simulated galaxies (see also Fig.~\ref{fig:sig_gas_tau05-5}), namely models G013, G025 with initial stellar masses of $10^{9.5}$~\Msun\ (top row), and G037 and G053 with initial stellar masses of $10^{10}$~\Msun\ (bottom row). The bottom panels in each row show the PDF normalised by the PDF at $\tau=1$ (NPDF). The faded grey curve shows a fit of a Log-normal to the PDF at $\tau=1$ of model G013 and serves as a reference in all panels.}
\label{fig:pdfnorm}
\end{figure}

To quantify further the state of the gas and the potential variation in clumpiness that appears in Fig.~\ref{fig:sig_gas_tau05-5}, we examine the mass-weighted gas density PDF that describes the relative importance of low- and high-density regions. In Fig.~\ref{fig:pdfnorm}, we show the evolution of the PDF inside a central cylindrical region of radius 1~kpc (0.5~\ls for those models) and height 2~kpc ($\pm 1$~kpc) as a function of time for four values of $\tau$ (as in Fig.~\ref{fig:sig_gas_tau05-5}). 

Overall, all models have gas densities from about $10^{-4}$ up to 10$^3$~cm$^{-3}$, initially with a profile roughly following a log-normal functional form. At $\tau = 1$ (once the bar has formed), all four models exhibit a similar PDF, following small adjustments relative to the PDF. For lower-stellar mass models G013 and G025, the overall shape of the PDF does not change significantly with time beyond the formation of the bar (top panels of Fig.~\ref{fig:pdfnorm}), except for a slight increase of high-density gas between $\tau =2$ and $\tau = 5$. This is consistent with the observed structure in the two-dimensional gas surface density map (Fig.~\ref{fig:sig_gas_tau05-5}): for those two models, the gas distribution at later times tends to be more centrally concentrated within the bar region, but its morphology still appears very clumpy and disorganised. 

In contrast with such a steady state, the central PDF of the higher stellar mass models G037 and G053 dramatically evolves at all scales after $\tau = 2$. At $\tau = 2$, the PDF develops a tail at high-density values reaching about $10^{3.5}$~cm$^{-3}$. This corresponds to the time when the bar-driven fuelling sets the seed of the central gas reservoir, a compact, dense and discy feature. Beyond $\tau = 2$, the shape of the PDF drastically changes. At $\tau$=5, the PDF is strongly skewed towards lower-density gas, with a bump in the gas density distribution at $\sim 10^{-4}$~cm$^{-3}$. The low-density bump is a direct consequence of the re-organisation of the gas structure within the bar, which creates a large contrast between central high-density regions and very gas-poor areas along the bar. 
The gas fuelling towards the inner disc is associated with the denser bar lanes, and those are surrounded by a depleted region between the bar ends and the inner gas disc. The abrupt cut at high densities certainly depends on resolution, as we would expect an extended power-law-like tail with higher-resolution simulations \citep[see e.g.][]{Renaud2013}. 
At $\tau=5$, when the inner gas and stellar inner discs are well settled, the PDFs in both G037 and G053 exhibit a steep profile with two main bumps: one around $10^2$~cm$^{-3}$ associated with the central gas reservoir, and one around $1$~cm$^{-3}$ associated with the bar gas lanes (Fig.~\ref{fig:pdfnorm}).

The contrast in evolution between the low-mass systems and the more massive ones is thus directly reflected in the PDF. Most importantly, it is challenging to discern whether a bar has formed in the lower-mass simulations by examining the central PDF, something already pointed out by \citet[][see their Figure~15]{Bland2024}.
\subsection{Dynamical tracers}
\begin{figure}[h!]
    \centering
    \includegraphics[width=1.\linewidth]{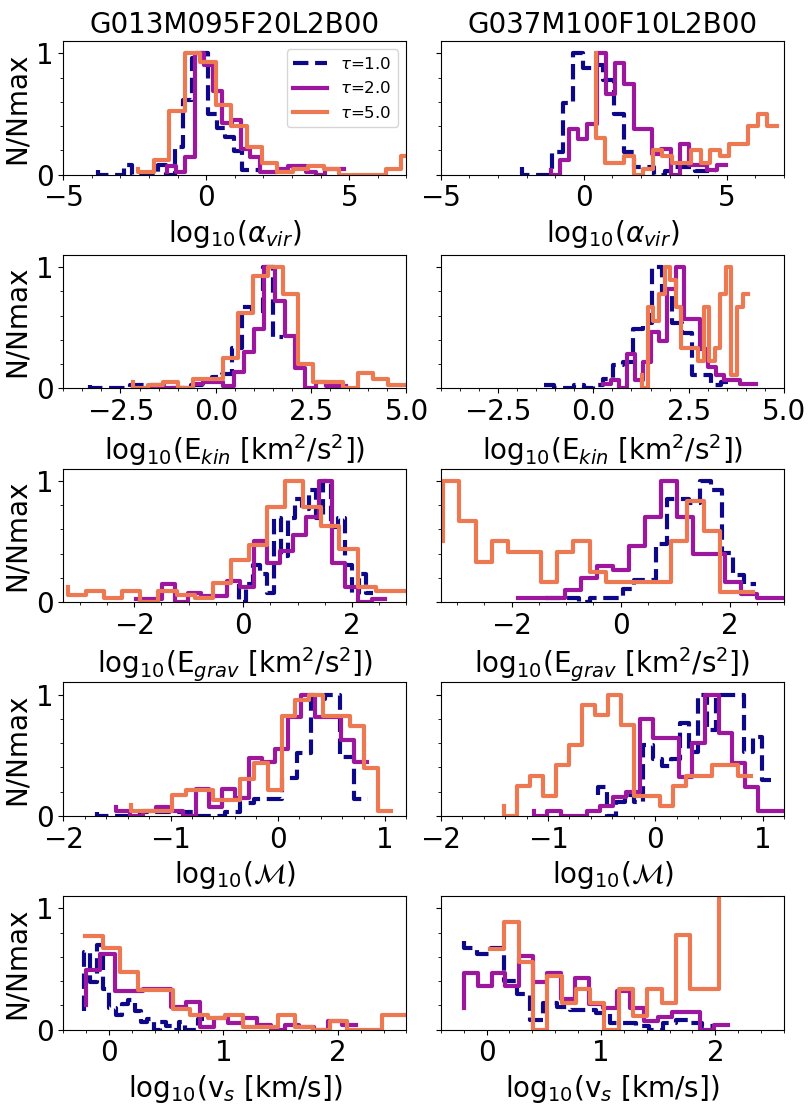}
    \caption{Evolution of the virial ($\alpha_{\rm vir}$), kinetic energy ($E_{\rm kin}$), potential energy ($E_{\rm grav}$), Mach number ($\cal M)$), and sound speed (v$_{\rm s}$) distributions for three values of the parameter $\tau$ for model G013 (left column) and G037 (right column).}
    \label{fig:vir_mach_G013_G037}
\end{figure}
\begin{figure}[h!]
    \centering
    \includegraphics[width=1.\linewidth]{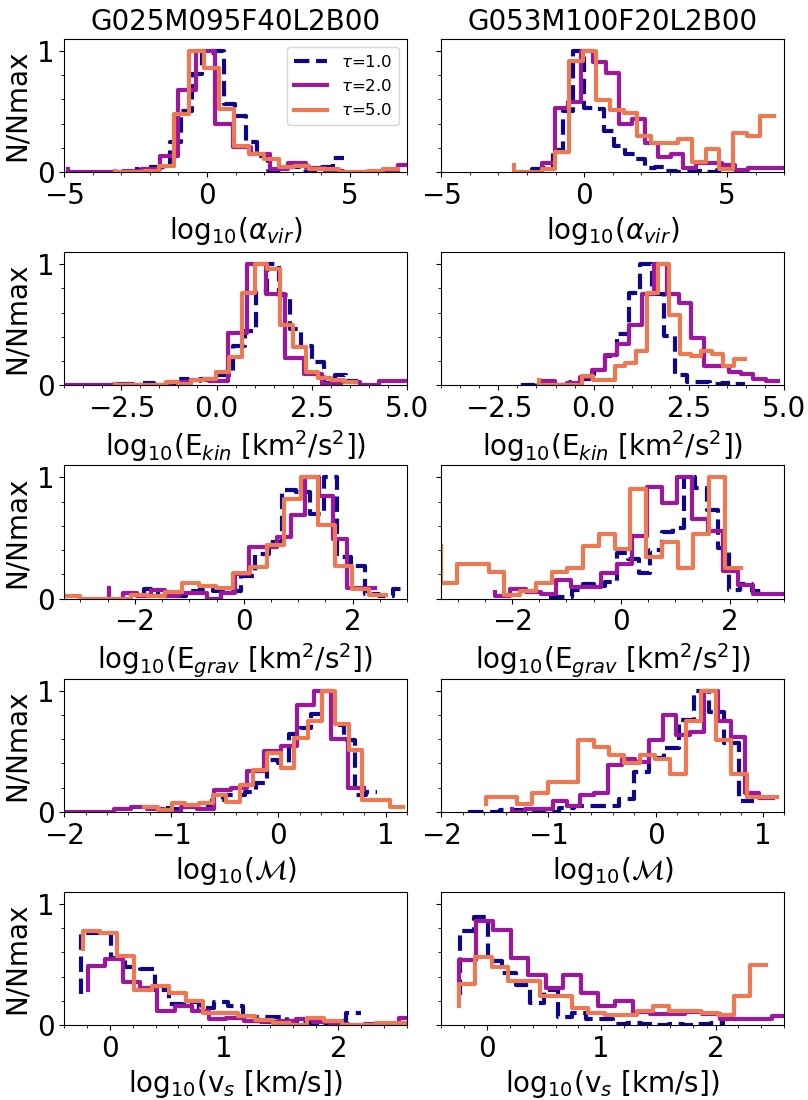}
    \caption{Same as in Fig.~\ref{fig:vir_mach_G013_G037}, but for models G025 and G053.}
    \label{fig:vir_mach_G025_G053}
\end{figure}

We next turn to dynamical tracers of the state of the gas in the bar region.
Figure~\ref{fig:vir_mach_G013_G037} illustrates the evolution over time of the distribution of the virial parameter ($\alpha_{\rm vir}$) and the Mach number $\cal M$. $\alpha_{\rm vir}$ addresses the ratio between the local kinetic and gravitational energies and is proportional to the square of the velocity dispersion and the inverse of the compactness of the clouds. The Mach number is the ratio between the velocity dispersion and the sound speed, which depends on the gas temperature. Those two quantities provide a quantitative description of the physical state of the star-forming gas and are often used as a relevant way to connect with observations \citep[e.g.][]{Sun2022}. To compute those distributions, we have used a centred box with a side length of 1~\ls\ (2~kpc) and a height of 2~kpc ($\pm 1$~kpc from the disc plane) and averaged the 1D velocity dispersion ($\sigma_{\rm 1D}$), sound speed ($v_{\rm s}$), potential energy ($E_{\rm grav}$), virial parameter ($\alpha_{\rm vir}$), and the Mach number ($\cal M$) of the smallest star-forming gas cells (12~pc) contained inside volumes of 100~pc$^{3}$. The typical number of events inside this volume varies between 5,000 and 150,000.

We first focus on the two relatively gas-poor models (concerning the SFMS), namely G013 (low-mass bin, with $10^{9.5}$~\Msun, and an initial gas fraction of 20\%) and G037 (higher-mass bin, with $10^{10}$~\Msun, and an initial gas fraction of 10\%). 
The distribution of the virial parameter and Mach number for G013 (left panels in Fig.~\ref{fig:vir_mach_G013_G037}) exhibit mild changes before and after the formation of the bar ($\tau=1$, blue dashed curve). 

The higher stellar mass model G037 (right panels in Fig.~\ref{fig:vir_mach_G013_G037}) shows a radically different behaviour. Both $\alpha_{\rm vir}$ and $\cal M$ distributions develop extended tails and secondary peaks after the formation of the bar (at higher and lower values, respectively). In Fig.~\ref{fig:vir_mach_G013_G037}, we also provide the respective numerators and denominators for $\alpha_{\rm vir}$ (resp. kinetic and gravitational energies) and $\cal M$ (resp. velocity dispersion and sound speed). For G037, the sound speed distribution exhibits a secondary peak at higher values that is directly related to an increase in the gas temperature. That temperature increase is caused by more intense stellar feedback associated with the emergence of a central disc, which is then reflected in the sub-sonic peak around values $\cal M \sim$ 0.25.

Those results are confirmed when examining models that lie closer to the SFMS galaxies, namely G025 (low-mass bin, with $10^{9.5}$~\Msun, and an initial gas fraction of 40\%) and G053 (higher-mass bin, with $10^{10}$~\Msun, and an initial gas fraction of 20\%). In Figure~\ref{fig:vir_mach_G025_G053}, the profiles in all panels associated with the lower mass galaxy simulation G025 (left column) are stable over time, while the tracers associated with the higher-mass model G053 evolve dramatically after the bar formation, in a qualitatively similar fashion as for G037.

The dynamical and physical status of the central star-forming regions does not seem to be affected by the formation of the bar in lower mass systems. This is quite a surprising result, but already expected from the constancy of the global PDF (see Fig.~\ref{fig:pdfnorm} and Sect.~\ref{sec:pdf}). In systems with masses above $10^{10}$~\Msun, the bar-driven evolution is dramatically different, leading to a contrasted set of low- and high-density regions (see again Fig.~\ref{fig:pdfnorm}), involving super-virial and sub-sonic gas.

As mentioned before, the drastic change in the evolution of such dynamical tracers could be related to the increase of the gas fraction in the low-mass systems. However, this does not seem to be the case, as a comparison between models G013 and G053 that share similar initial conditions (only with a mass scaling) and the same gas fraction of 20\% shows consistent trends.
\section{The balance between stellar feedback and gravity}\label{sec:feedback_grav}

In this section, we probe the physical origin of the observed change of regime traced via a variation of local values of $\cal M$ and $\alpha_{\rm vir}$ (Sect.~\ref{sec:tracers}) by studying the balance between stellar feedback and the gravitational potential, and the ratio between the ejected gas mass and the SFR (mass loading factor). Those two tracers allow us to understand the clumpy gas distribution in our lower-mass models and the emergence of an inner gas disc in higher stellar mass models.
\subsection{Feedback versus gravity}
\label{sec:feedgrav}
\begin{figure*}[h!]
\centering
\includegraphics[scale=0.5]{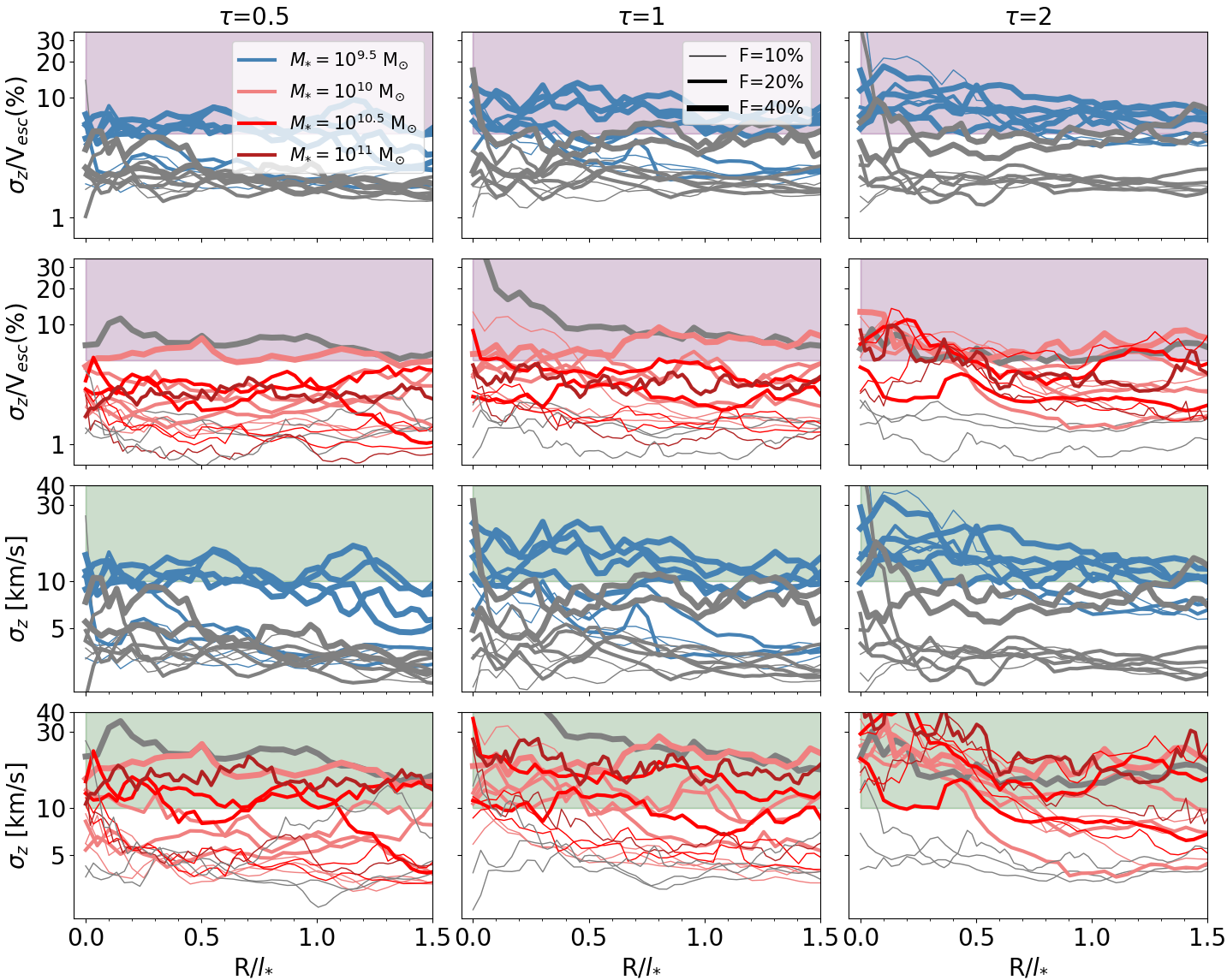}
\caption{Evolution (from left to right) of the vertical velocity dispersion $\sigma_z$ (two bottom rows) and $\sigma_z$ normalised by the local escape velocity (two top rows). The blue (resp. red) curves represent models with an initial stellar mass of $10^{9.5}$~\Msun (resp. above or equal to $10^{10}$~\Msun, see inset in  top left panel). The grey lines are models that do not display a bar at the end of the simulation. The line thickness corresponds to the gas fraction (see inset in the top middle  panel). The green shaded areas show vertical velocity dispersions ranging from 10 to 40 \kms and the purple shaded areas show a percentage above 5\%.}
\label{fig:feedback_potential}
\end{figure*}
Our simulations include stellar feedback in the form of supernovae and stellar winds. These processes inject energy and momentum into the surrounding ISM, leading to an increase of the local gas velocity dispersion $\sigma$ \citep[e.g.][]{MaclowKlessen2004,Padoan2017,Agertz2009}. When vigorous enough, feedback can drive galactic-scale outflows and expel gas far beyond the stellar disc \citep[for reviews, see e.g.][]{Veilleux2005, Rupke2018}, a process that is well studied in numerical simulations \citep[e.g.][]{Muratov2015,Nelson2019,Andersson2023}. One direct tracer of the impact of feedback is the variation in the velocity dispersion, and more specifically, its vertical component \sigz. Such outflowing gas is pulled back by the gravitational potential of the disc that may dominate (or not) the local energy budget, preventing the gas (or not) from escaping towards large vertical distances.

To quantify the relative effect of feedback in our simulation, we measured the vertical velocity dispersion \sigrel\ normalised by the local escape velocity \vesc, the latter being directly connected with the depth of the local total gravitational potential for all simulations in our extended set. The evolution over time of the dispersion and its normalised version is shown in Fig.~\ref{fig:feedback_potential}.
We emphasise that the escape velocity profiles do not significantly change with time. Hence, most of the evolution observed in the normalised dispersion values can be linked back to an evolution in the velocity dispersion itself.

Remarkably, all unbarred systems (but G069) show consistent and flat vertical velocity dispersion profiles with values between $\sim 3$ and 8~\kms, with an increase for a few systems restricted to the central region ($R < 0.5$~\ls). The only exception is model G069: as already noted (see Sect.~\ref{sec:illustration}), that model with an initial gas fraction of 40\% develops a bar which subsequently gets dissolved. Its vertical dispersion profile is consistent with other gas-rich systems, and its dispersion settles back with a flat and lower value profile after the bar dissolution (grey thick line in the lower right panel at $\tau=2$ in Fig.~\ref{fig:feedback_potential}).

At fixed gas fraction, the vertical dispersion is higher at $\tau = 0.5$ (before the formation of the bar) in higher stellar mass systems, as expected. This trend starts to disappear for all barred systems  (red and blue lines in Fig.~\ref{fig:feedback_potential}) after the formation of the bar with a distribution of $\sigma_z$ values spread between 10 and 20~\kms\ at $\tau = 2$. For those same systems, we also witness a trend that higher gas fraction systems have higher values of \sigz: this is expected considering that a higher gas fraction relates to a higher SF rate, hence a relatively stronger role for stellar feedback, as predicted by  \citet{Ostriker2010}, among others. By $\tau=2$, all barred systems with at least 20\% of gas exhibit \sigz\ values between 6 and 20~\kms\ with peaks up to 30~\kms. This again contrasts with the sustained and lower \sigz\ values of unbarred systems, as mentioned. The bar thus seems to have a prominent role in establishing and sustaining a higher vertical velocity dispersion. We finally notice that barred discs tend to have higher dispersion values towards the centre at $\tau = 2$.

The velocity dispersion profiles normalised by local escape velocities are presented in the top rows of Fig.~\ref{fig:feedback_potential} with different colours depicting different stellar masses (from blue for lower mass to dark red for higher mass). At $\tau = 0.5$, most profiles are relatively flat and spread over a range from about 1 to 8\%, again with a tendency for systems with higher gas fractions to have higher normalised \sigz.  By the time the bar has formed ($\tau = 1$), we start seeing a relative increase of the lower stellar mass systems, and at $\tau = 2$ there seems to be a clear stellar mass segregation: discs with $10^{9.5}$~\Msun\ stellar mass have normalised \sigz\ all above 5\% and up to 11\% while higher-mass systems are spread in a lower range between 2 and 6\%. As expected from the above-mentioned trends, most unbarred simulations have low values (between 1 and 2\%), and this stays at all times.

In summary, we used the vertical dispersion as a tracer of feedback:
\begin{itemize}
    \item \sigz\ seems to increase with time for all barred systems, while unbarred discs keep relatively low and flat dispersion profiles (with \sigz\ around 5~\kms);
    \item Higher gas fraction systems tend to have higher \sigz\ values, but this trend disappears after the formation of the bar.
    \item \sigz\ profiles at $\tau = 2$ for barred systems seem to have no significant dependence on stellar mass, while the normalised profiles exhibit a clear segregation with stellar mass, with lower (resp. higher) stellar mass systems having significantly higher (resp. lower) normalised \sigz.
\end{itemize}
Such findings emphasise the role of the bar in establishing a turbulent gaseous medium in the central region of those galaxies, as well as the induced and abrupt change in the relative importance between stellar-driven feedback and the local gravitational potential.

\subsection{The crucial role of supernovae in preventing the formation of an inner gas disc}
\label{sec:sn}

As shown in the previous section and Fig.~\ref{fig:feedback_potential} both the formation of a bar and the depth of the gravitational potential have a clear impact on the physical status of the ISM in our simulations. It suggests that bars in lower-mass systems have a stronger impact due to both the shallower depth of the gravitational potential (as compared with higher stellar mass discs) and the observed higher gas fraction of main-sequence SF. We now bring evidence that stellar feedback, as implemented via sub-grid recipes in our simulations, is the driver of such a trend with the bar in the lowest mass systems. We thus run a test case using model G001 (a low stellar mass model with a gas fraction of 10\%) as a reference, where we turn off the feedback from supernovae (model named G001-NOSN). In that simulation, only the stellar feedback associated with stellar winds is kept active. 

\begin{figure*}[h]
\centering
\includegraphics[width=0.9\textwidth]{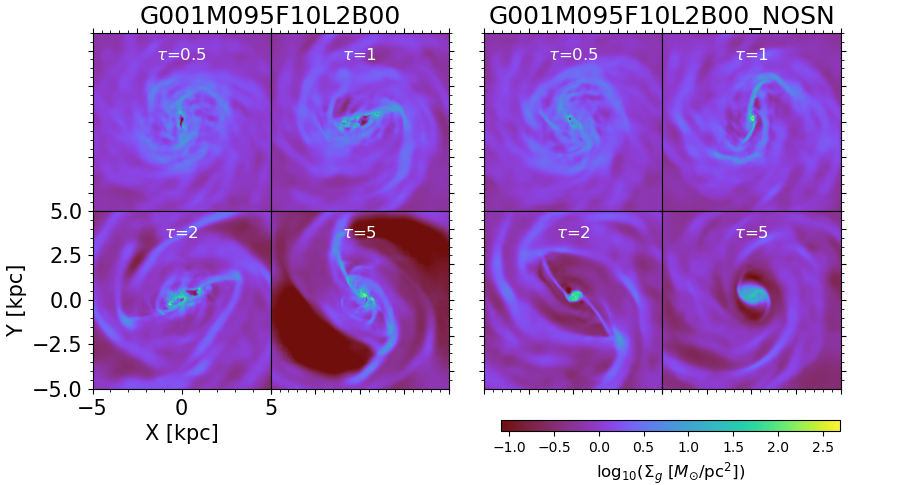}
\caption{Surface density map of gas of model G001 and G001-NOSN. NOSN stands for the run without feedback from supernovae and only includes feedback from stellar winds. Each panel shows one model at different values of the parameter $\tau$.}
\label{fig:nosn}
\end{figure*}

\begin{figure*}[h]
\centering
\includegraphics[width=0.92\textwidth]{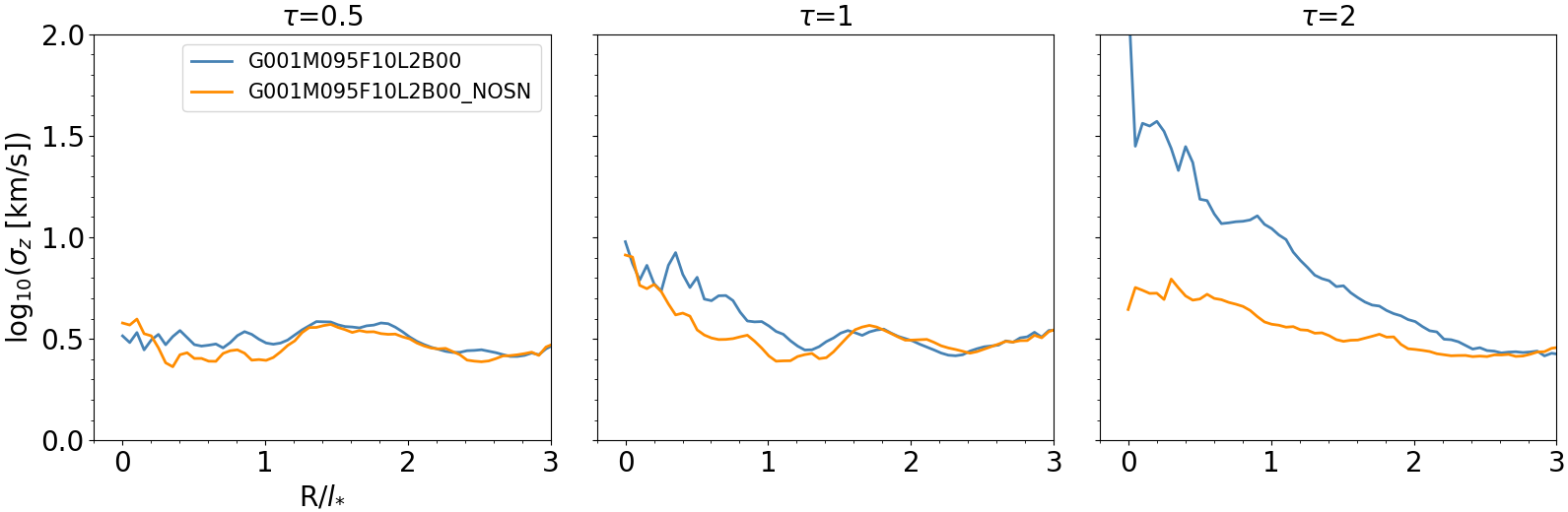}
\caption{Same as bottom panels of Fig.\ref{fig:feedback_potential}, but for model G001 (blue curve) and G001-NOSN (orange curve).}
\label{fig:velocity_dispersion_test_models}
\end{figure*}

Figure~\ref{fig:nosn} illustrates the evolution of model G001 with (left panel) and without (right panel) supernovae feedback. We observe that the onset of a central disc is already visible in the model without feedback (G001-NOSN) at $\tau = 1$, and further develops into an extended central gas reservoir at $\tau = 5$. The impact of supernova feedback on the gas velocity dispersion is directly illustrated in Fig.~\ref{fig:velocity_dispersion_test_models} where we present the $\sigma_z$ radial profile for both G0001 models (with and without supernova feedback). As expected, model G001-NOSN has a low and rather constant vertical velocity dispersion ($\sigma_z \sim 3$~\kms). Even though G001-NOSN develops a strong bar, its $\sigma_z$ profile does not vary much with time and is consistent with those observed for all our unbarred models in Fig.~\ref{fig:feedback_potential}.
\begin{figure*}[h!]
    \centering
    \includegraphics[width=0.92\textwidth]{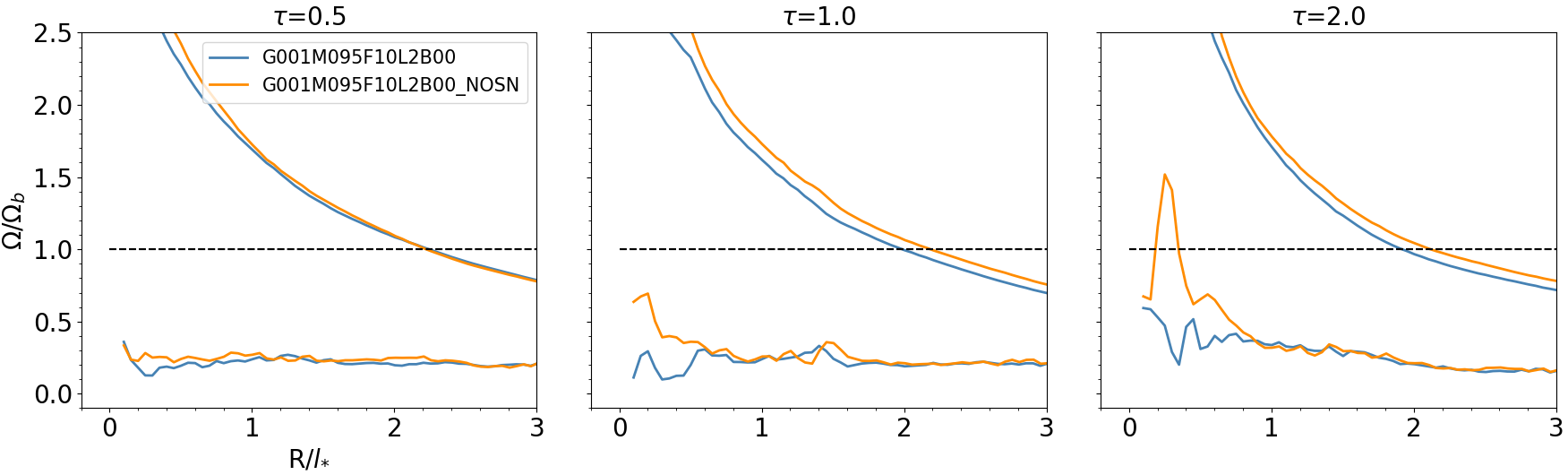}
    \caption{Same as in Fig.~\ref{fig:ILR}, but for model G001 (blue curve) and G001-NOSN (orange curve).}
    \label{fig:ilr_test_models}
\end{figure*}

The emergence of a central gas reservoir associated with a young stellar disc is further consolidated by the observed evolution of the frequency profiles, as shown in Fig.~\ref{fig:ilr_test_models}: in contrast with the G001 models, G001-NOSN exhibits a broad central increase of the $\Omega - \kappa / 2$ at $\tau = 1$ and develops a clear ILR as observed in the profile at $\tau = 2$. Interestingly, model G001, that includes supernova feedback, exhibits a mild increase of its $\Omega - \kappa / 2$ profile towards the end of the simulation ($\tau = 5$), suggesting that it may manage to build a bona fide central gas reservoir but on a much longer timescale than more massive systems (see the discussion in Sect.~\ref{sec:conclusion}).

Running test simulations such as G001-NOSN for all our simulations is beyond the scope of the present paper. Still, G001-NOSN provides strong evidence that stellar feedback, and more specifically supernovae feedback, is the main driver in our simulations for the change of regime between low-mass and higher-mass systems, and prevents the emergence of a central gas reservoir in galaxies with shallow gravitational potentials.

\section{Discussion and conclusion} 
\label{sec:conclusion}

\subsection{A gas reservoir chronicle}
\begin{figure*}[h!]
\centering
\includegraphics[width=0.9\textwidth]{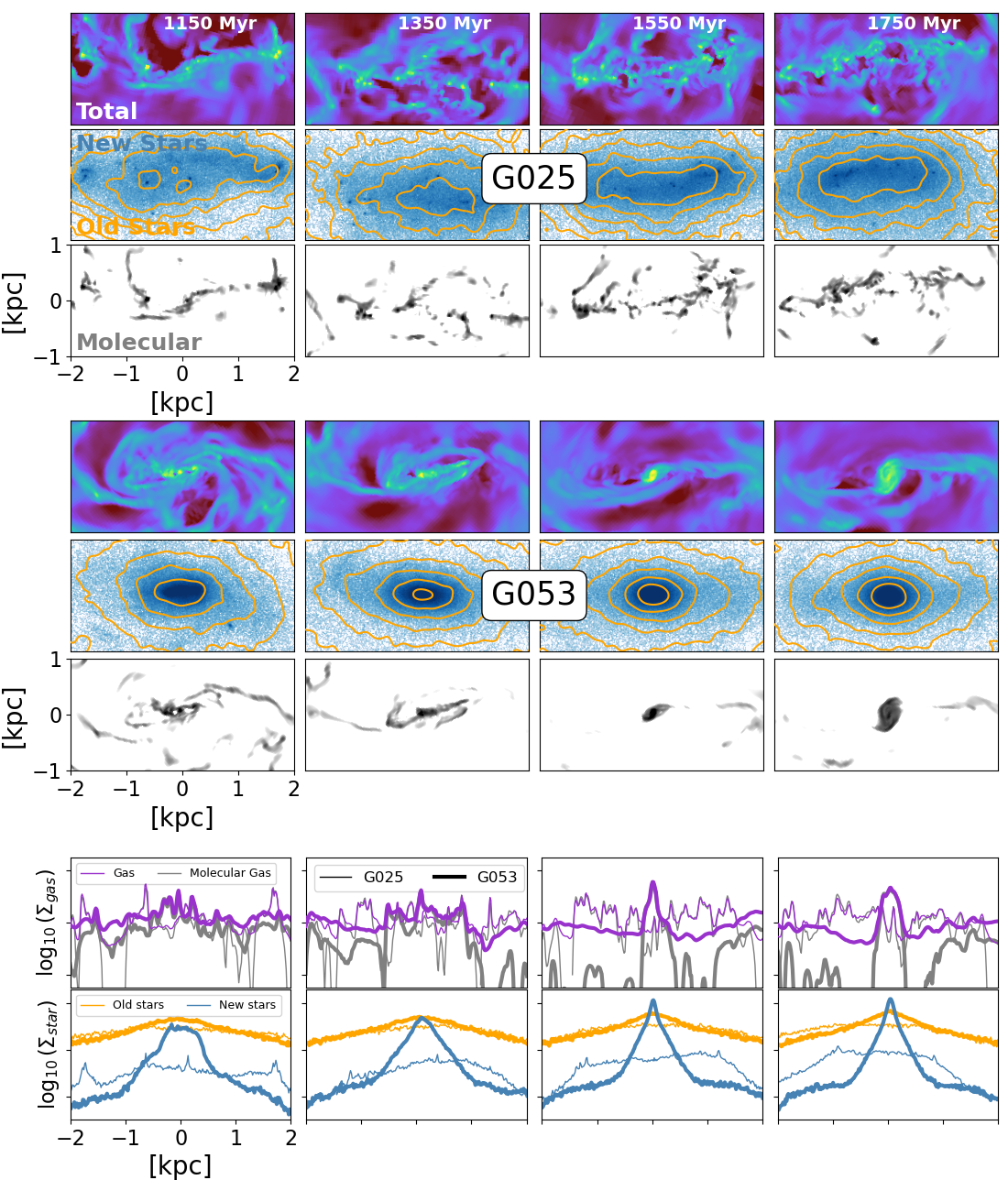}
\caption{Time evolution of simulations G025 and G053,  a low and intermediate stellar mass system ($10^{9.5}$ and $10^{10}$~\Msun, respectively) in the central 2~kpc. Four running times are presented (from left to right, first six panels from top), with steps of 200~Myr, with a seed gas reservoir forming in G053 around $t=1400$~Myr (between the two middle panels). Maps with a field of view of 2~kpc $\times$ 1~kpc of the total gas surface density, the mass of new stars (and old stars, as orange contours) and molecular gas density are shown as maps in the first to third rows for G025, and fourth to sixth rows for G053. In all maps, the bar is aligned with the x-axis. The two bottom rows show the radial profiles, averaged over the y-axis ($\pm 1$~kpc), first with the gas profiles (total and molecular), and then the stellar density profiles (bottom row; old and new stars). }
\label{fig:seedform}
\end{figure*}
In the previous sections, we have provided evidence that lower mass simulations of disc galaxies exhibit a different evolution in their gas redistribution driven by a stellar bar. We have identified that stellar-driven feedback is a key factor in preventing a coherent structure from emerging. In higher-mass systems, the gas reservoir is built while new stars have already started to accumulate within the central region of the bar (see e.g. Fig.~\ref{fig:sig_gas_tau05-5}). 

The emergence of those structures is emphasised in Fig.~\ref{fig:seedform} where we present a time sequence encapsulating the early formation of the gas reservoir for two reference simulated systems that lie on the SFMS, namely G025 ($10^{9.5}$~\Msun) and G053 ($10^{10}$~\Msun). Only the G053 higher-mass run forms a well-structured gas flow and central gas disc structure in the first 3 Gyr (at a running time of about $t=1400$~Myr). Over 600~Myr, the lower mass G025 run shows very little evolution in the total gas, molecular gas and stellar distribution. For G025, high gas density clumps are observed mostly travelling along the bar, significantly disturbed by continuous injection of feedback energy, and its associated one-dimensional gas density profile is flat. The case of G053 is radically different, with a very strong time evolution: while the averaged gas density profile along the bar axis resembles the G025 one in the first two panels, a clear peak forms after 1400~Myr (last two panels) surrounded by a region of depressed gas density. This is reminiscent of the observed signature from the bar-driven organised gas flow seen in the profiles presented in Fig.~\ref{fig:surf_tau05-5}. This is particularly visible in the molecular gas distribution and profile for G053, where the bar region is almost entirely devoid of gas except for the emerging central $\sim 200$~pc gas reservoir. It is also remarkable to follow how new stars preferentially accumulate in a growing and well-defined central exponential disc (with a scale length of about 100~pc when projected edge-on).

\subsection{The importance of timescales}
\label{sec:timescales}

The emergence of a central gas reservoir appears to occur at early times for models with larger stellar masses only. This is illustrated in the formation of an ILR that is associated with a central (stellar and gas) mass concentration (Fig.~\ref{fig:ilr_test_models}). However, in Fig.~\ref{fig:nosn}, we observe a late onset of a more organised fueling flow for a lower-mass system (G001), which is also reflected in the slight central increase in the $\Omega - \kappa / 2$ radial profile of Fig.~\ref{fig:ilr_test_models}. This suggests that the absence of a central gas reservoir at $\tau = 5$ in lower mass galaxies can be the consequence of a longer timescale needed in these smaller systems to acquire gas via a coherent, inward flow rather than of a stellar mass criterion alone. 

To test this hypothesis, we have run an additional low stellar mass case using the same conditions as for the G001 model, but this time running up to 7~Gyr, corresponding roughly to $\tau = 10$. Up to a time $t \sim 5.5$~Gyr, the gas distribution resembles other low-mass simulations, with a rather disorganised, clumpy appearance. At $t \sim 5.7$~Gyr, the gas flow tends to a more ordered organisation, and by $t = 5.85$~Gyr (i.e. $\tau >8$), the seed of a central gas disc appears. If we compare such a timescale with other larger-mass models at the same gas fraction, it is more than four times faster for model G037 (with an initial $10^{10}$~\Msun\ stellar mass) to reach a similar state (both in absolute terms of megayears and in terms of the corresponding $\tau$). It is therefore likely that the observed transition between a stellar mass of $10^{9.5}$ and $10^{10}$~\Msun\ also stems from a continuous varying timescale for the formation of a reservoir that depends in turn on the relative importance of feedback and gravity (see Sect.~\ref{sec:feedgrav}).  

\begin{figure*}[h!]
\centering
\includegraphics[width=0.9\textwidth]{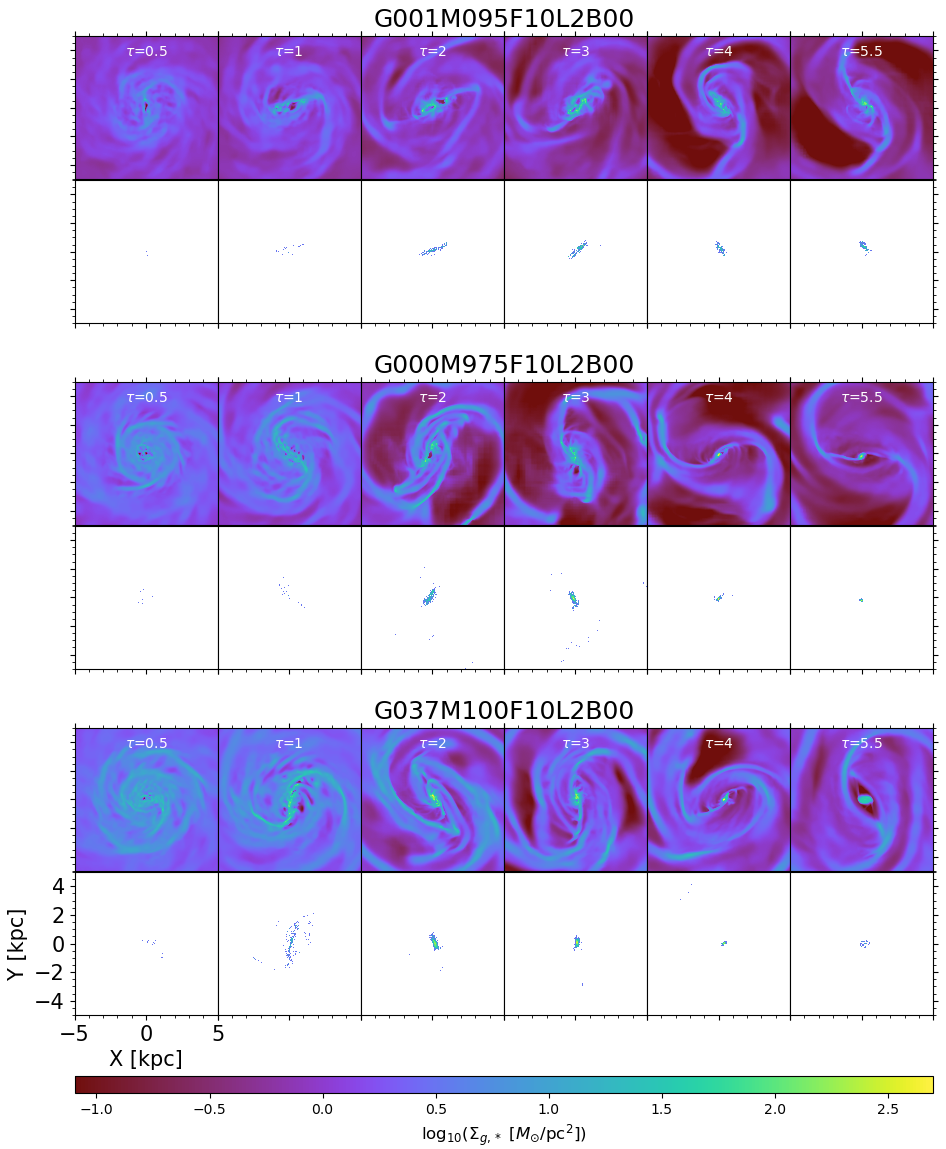}
\caption{Illustration of the morphological transition using models G001 (top panel), G000M975F10L2B00 (middle panel), and G037 (bottom panel), corresponding to a low-, low-intermediate, and high stellar mass model, respectively. In each panel we show the evolution of the surface density map of gas at six different values of $\tau$ (0.5, 1, 2, 4, 5, and 6).}
\label{fig:phase_transition}
\end{figure*}

To probe such a transition further, we performed a simulation using an intermediate-stellar mass, namely $10^{9.75}$~\Msun\ labelled G000M975F10L2B00. The resulting gas evolution is shown in Fig.~\ref{fig:phase_transition} where we show the comparative evolution of our two reference models G001 (top panel; $10^{9.5}$~\Msun) and G037 (bottom panel; $10^{10}$~\Msun). We observe the formation of the gas reservoir and then the gas reservoir after $\tau=4$ for the intermediate-mass models, while it forms at $\tau \sim 3$ and $\tau \sim 8$ for the higher-mass and lower-mass ones, respectively. The radial extent of the gas reservoir at $\tau=5.5$ for this intermediate model is significantly larger than for the high stellar mass model. This trend stands when using absolute times (formation time of the seed at about 5.9, 2.3 and 1.7~Gyr, respectively) or the time it takes after the formation of the bar (5.1, 1.8, and 1.2~Gyr, or 12, 5, and 5 bar rotations, respectively). This strongly suggests that the formation timescale of the gas reservoir after the bar has formed depends on the galaxy's stellar mass and that this timescale becomes significantly larger when transitioning from discs with $10^{10}$ down to $10^{9.5}$~\Msun, preventing a compact gas (and stellar) disc to form within a few Gyr after the formation of the bar in the lower mass systems. The limited set of values for the formation timescales of the inner gas reservoir we obtained suggests a superlinear dependence on stellar mass. From the time between $\tau=1$ (bar formation) and the advent of the central mass seed, we estimate a timescale that is proportional to M$^{-3/2}$. While the true dependence needs to be confirmed with additional experiments, it has to be significantly steeper than an inverse linear ($1 / \rm{M}$) trend to imprint such an abrupt increase towards lower masses.

\subsection{An evolutionary perspective}
\label{sec:scenario}

Following the results from Section~\ref{sec:feedback_grav}, we expect a continuous change in the balance between stellar feedback and the depth of the gravitational potential when stellar mass varies. At fixed gas fraction, as host galaxy stellar masses decrease in the range $10^{9.5} - 10^{10}$~\Msun, the gas in each system re-organised by the bar and fuelled towards the central region becomes more impacted by supernovae feedback and thus tends to be less prone to forming a central mass concentration. The amplitude of such a change, when transitioning between $10^{10}$ down to $10^{9.5}$~\Msun, leads to the apparent dichotomy in SF regimes and structuring of the bar region between the lower- and higher-mass ends. In this section, we provide a schematic and qualitative scenario for the evolution of star-forming main-sequence galaxies when a bar forms, which may help explain this abrupt change of timescale.

When a bar forms in a disc system, the non-axisymmetric and rotating part of the potential establishes a new orbital skeleton that drives a complex set of trajectories for the stars and the (dissipative) gas \citep[see e.g.][]{Atha1992bar, Ceverino2007}. Without a mass concentration, there is initially no inner resonance (ILR), the gas is channelled along radial orbits towards the centre \citep[see Sect.~3.3 of ][]{Atha1992gas}, SF can be induced in the region of convergence of the two incoming bar lanes.

For relatively massive stellar systems, the feedback energy does not overcome the torque exerted by the potential from the gravitational potential and does not manage to disrupt the gas flow significantly. This leads to an organised flow from which a central area naturally emerges at the region where opposite bar lanes meet. Mass starts to accumulate in the central region of the bar as the inflowing gas contributes to forming stars: a central stellar disc is formed on a timescale of a couple of bar rotations. Such a disc further helps to focus subsequent fuelling and central star-forming episodes, finally leading to the onset of an ILR.

In lower mass systems, the stellar-driven feedback is sufficient to disrupt the organised gas flow within the bar. Instead of a focused set of bar lanes, it leads to an elongated but thick three-dimensional clumpy structure (see Fig.~\ref{fig:seedform}): SF proceeds all along the bar region with star-forming clumps being distributed along the bar but spread over a wide range perpendicularly to its major-axis. As illustrated in the top panels of Fig.~\ref{fig:seedform}, all portions of the bar traced by either gas or new stars look similarly chaotic: there is no well-defined centre. 

The typical velocities of the gas contained in the central kiloparsec in our simulations with initial stellar masses of $10^{9.5}$~\Msun\ are in the range from 20 to 30~\kms, hence commensurate with the dispersion induced by stellar-driven feedback that we show in Fig.~\ref{fig:feedback_potential} to be mostly mass-independent. Figure~\ref{fig:feedback_potential} also indicates that above a stellar mass of $10^{10}$~\Msun\, the contribution of feedback to the dispersion becomes negligible in relative terms. In light of our study of the simulations in Sect.~\ref{sec:timescales}, we find that the result is a clear impact on the characteristic timescale at which the central reservoir is built by the bar. 

The simulations suggest a characteristic timescale proportional to $\rm{M}^{-3/2}$, or a slowed rate at low stellar mass in contrast with a simple inverse scaling with stellar mass, for example. The mass dependence of this formation timescale potentially connects to the balance between feedback and the local potential and, more specifically, between the inflow rate induced by the bar and the ability of supernovae feedback to disrupt such a flow. We expect this mass scaling to reflect the tradeoff between the rate at which the bar drives gas inwards and the ability of supernovae feedback to disrupt such a flow. The variables that would lead to such a mass scaling would then include the inflow rate associated with the bar, the energy deposited by supernovae feedback and how it couples with the local gas.
The relation between these quantities and stellar mass follows from scaling relations, i.e. SFR and gas fraction variations on the main sequence, and the characteristic radius of the bar. The fact that the transition between organised and disorganised flows occurs when the feedback-driven dispersion becomes significant (Fig.~\ref{fig:feedback_potential}) provides a compelling basis for a predictive model, something that is beyond the scope of the present paper. 

The mass dependence of the growth timescale found in this work, together with the evidence that the central gas velocity dispersions originate with feedback in galaxies below $10^{9.5}$~\Msun, support a picture to explain the evolutionary change we witness in observations: if bars are young enough in low-mass galaxies, they would never have time to meet the conditions to form a central gas reservoir. Those conditions would only be met when the specific SF rate drops low enough for the gas flow to become organised and facilitate the build-up of a central seed mass. In this light, it may be relevant that galaxies on the SFMS exhibit a significant increase of the gas fraction with decreasing mass below a stellar mass of $10^{10}$~\Msun, potentially signifying a further lengthening of the time it takes for an isolated low-mass disc to form a central mass seed. This is, of course, a speculative evolutionary scenario that should be further examined and tested with a larger set of simulations. Such a desirable set of simulations would require long running times and hence an investment of CPU time well beyond the realm of the present paper. Resolved observations obtained via the PHANGS campaign will also provide further insights, potentially in ways that were not yet explored.

\subsection{Summary}
For this work, we have used our extended grid of simulations (35 models) based on 4 galactic parameters (i.e. stellar mass, gas fraction, scale length of the disc of stars, and the bulge mass fraction) coming from the PHANGS sample to investigate the main factors involved in the emergence of central gas reservoirs (central molecular discs or rings). We have noticed the presence of central gas reservoirs in our simulations and observed galaxies above a certain stellar-mass threshold corresponding approximately to $10^{10}$ \Msun, setting the low and high stellar mass regimes. Those gas reservoirs are observed in the surface density maps and in the specific distribution of SF across the disc. When a stellar bar has formed, we observe that the star-forming regions remain inside the bar in the low stellar mass regime. In the high stellar mass regime, we observe a redistribution of the star-forming regions into a central gas reservoir and near the bar ends. We observe the formation of a central gas reservoir for all high stellar mass models of Table~\ref{Tab:Grid_35} exhibiting a stellar bar \citep[but for model G178, see][for more details]{Verwilghen_2024}.

The presence of a gas reservoir in high stellar mass models creates a depletion of gas between the interior of the stellar bar and the gas reservoir. This gas depletion in the high stellar mass models leads to a dip in the gas surface density profiles that we do not observe in the low stellar mass models. In addition to this gas depletion, we have also studied the physical properties of the star-forming gas via the PDF, the distribution of the virial parameter, and the Mach number. We have shown that those physical properties remain unchanged over time in the low stellar mass regime, but vary significantly as a function of time in the high stellar mass regime. We highlight here that this gas depletion causes a dip in the gas surface density profiles, and the distribution of the mentioned physical parameters could be compared with real galaxy data in future work (Emsellem et al. in prep. Neumann et al. in prep.).

We computed the circular velocity coming from the gravitational potential, derived the angular frequency, and the bar pattern speed to check whether an ILR has formed in our barred models. We found that in the high stellar mass regime, an ILR has emerged, which illustrates the formation of central gas reservoirs. We did not identify any ILR in the low stellar mass regime.

We have investigated the role played by the stellar feedback and compared it with the gravitational potential of our different models. We have found that the variation of the gas vertical velocity dispersion, mostly induced by SN feedback, has the same order of magnitude (15-30 \kms) for all our barred models. The ratio between the vertical velocity dispersion and the escape velocity is significantly higher (10-15\%) in the low stellar mass regime compared with the high stellar mass one ($\sim$ 5\%).

The change of regime is not dictated by an abrupt stellar mass threshold around $10^{10}$ \Msun, but rather corresponds to a significant and continuous increase of the timescale for the formation of a central mass concentration between stellar mass $10^{9.5}$ and $10^{10}$ \Msun. This transition follows the relative balance between the SN feedback and the local gravitational potential. In more massive systems, it allows an organised flow that builds a central mass concentration, ultimately leading to the emergence of an ILR. This picture is supported by a lower mass test model (G001\_NOSN) that forms a gas reservoir when SN feedback is switched off.

Our results shed new light on the observed change of structures witnessed in observations \citep[see e.g.][Gleis et al., in preparation]{FraserMcKelvie2020,Stuber+2023}.
Further work is now needed to consolidate this view and better understand the (potentially superlinear) dependence of the gas reservoir formation timescale on the galactic mass and potentially on other parameters (e.g. gas fraction). A toy model focusing on the balance between stellar-driven feedback and the gravitational potential may be a way to more directly connect our findings with the observed morphologies of nearby star-forming main-sequence galaxies. It would also help quantify the potentially significant impact of such differential evolutionary tracks on the more global evolution of galactic discs.
\begin{acknowledgements}
The authors gratefully acknowledge the Gauss Centre for Supercomputing e.V. (www.gauss-centre.eu) for funding this project by providing computing time on the GCS Supercomputer SuperMUC (Project pn57wu) at Leibniz Supercomputing Centre (www.lrz.de).
PV acknowledges support from the Excellence Cluster ORIGINS funded by the Deutsche Forschungsgemeinschaft (German Research Foundation) under Germany’s Excellence Strategy – EXC-2094–390783311. The simulations in this paper have been carried out on the computing facilities of the Computational Center for Particle and Astrophysics (C2PAP). We are grateful for the support by Alexey Krukau and Margarita Petkova through C2PAP. 
This work is based on observations made with the NASA/ESA/CSA JWST and we specifically acknowledge the use of data from JWST Programme GO 3707 (PI Leroy).
MV is supported by the Italian Research Center on High Performance Computing, Big Data and Quantum Computing (ICSC), project funded by European Union - NextGenerationEU - and National Recovery and Resilience Plan (NRRP) - Mission 4 Component 2, within the activities of Spoke 3, Astrophysics and Cosmos Observations, and by the INFN Indark Grant.
RSK and SCOG acknowledge financial support from the European Research Council via the ERC Synergy Grant ``ECOGAL'' (project ID 855130),  from the German Excellence Strategy via the Heidelberg Cluster of Excellence (EXC 2181 - 390900948) ``STRUCTURES'', and from the German Ministry for Economic Affairs and Climate Action in project ``MAINN'' (funding ID 50OO2206). 
RSK is grateful for computing resources provided by the Ministry of Science, Research and the Arts (MWK) of the State of Baden-W\"{u}rttemberg through bwHPC and the German Science Foundation (DFG) through grants INST 35/1134-1 FUGG and 35/1597-1 FUGG, and also for data storage at SDS@hd funded through grants INST 35/1314-1 FUGG and INST 35/1503-1 FUGG. 
RSK also thanks the Harvard-Smithsonian Center for Astrophysics and the Radcliffe Institute for Advanced Studies for their hospitality during his sabbatical, and the 2024/25 Class of Radcliffe Fellows for highly interesting and stimulating discussions. 
OA acknowledges support from the Knut and Alice Wallenberg Foundation, the Swedish Research Council (grant 2019-04659), the Swedish National Space Agency (SNSA Dnr 2023-00164), and the LMK foundation.
\end{acknowledgements}
\bibliographystyle{aa}
\bibliography{main}
\begin{appendix} 
\section{The gas and SF distribution in non-barred systems}
\label{sec:appa}

The models we mainly use as illustrations in this work are the ones forming bars (Y, Tab.~\ref{Tab:Grid_35}) because they highlight the two stellar-mass regimes. In this section, we show four models which never form a stellar bar (i.e. \A\ $\leq$ 0.2) or do not keep it until the end of the run ($\sim$ 3000 Myr) to compare them with the four barred models in Fig.~\ref{fig:sig_gas_tau05-5}.

\begin{figure}[h!]
    \centering
    \includegraphics[width=\linewidth]{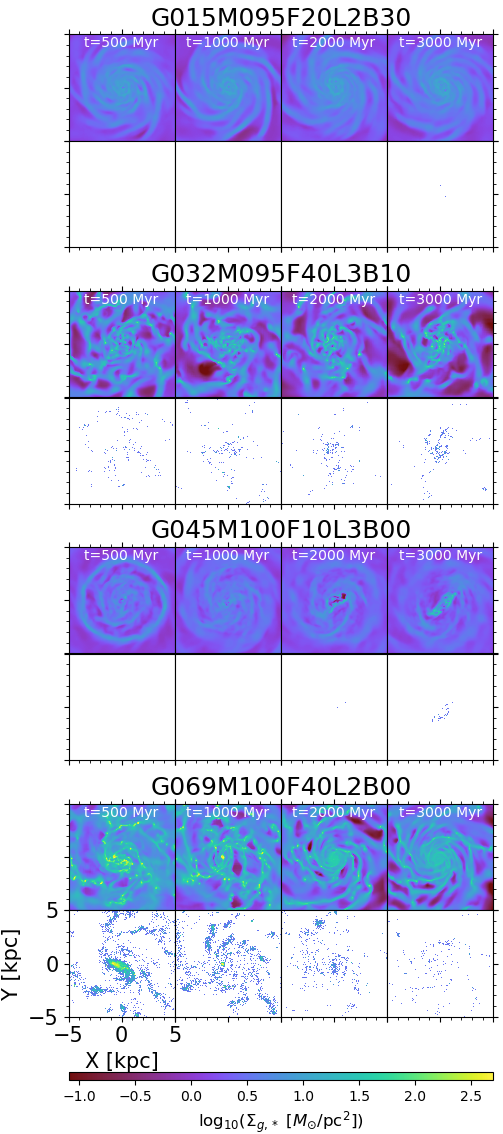}
    \caption{Same as in Fig.~\ref{fig:sig_gas_tau05-5}, but for four non-barred models: G015, G032, G045, and G069.}
    \label{fig:gas_map_nobar}
\end{figure}

Figure.~\ref{fig:gas_map_nobar} illustrates the gas and newly formed star density maps at four different time steps (i.e. 500, 1000, 2000, and 3000 Myr) of four of our models. In these figures, we show the two lower-stellar mass models G015 and G032 ($10^{9.5}$ \Msun, top panels) and two higher stellar mass models G045 and G069 ($10^{10}$ \Msun, bottom panels). In the two top panels, the lower gas fraction model G015 (F=20\%) displays only a spiral structure that does not significantly evolve with time. The higher gas fraction model G032 (F=40\%) also shows some spiral structures, but the ISM is more turbulent because of the higher SFR due to a higher gas fraction. The two bottom panels show models with gas fractions of F=10 (G045) and 40\% (G069). Model G045 also shows a spiral pattern but starts to develop a bar-like structure between 2 and 3 Gyr. We do not consider this structure as a bar because its \A\ Fourier coefficient stays below the value of 0.2. We observe in Fig.~\ref{fig:gas_map_nobar} that the SFR of this model remains low until t=2000 Myr. 
\section{Interaction with a massive stellar cluster}
\label{sec:appb}

\begin{figure*}[h!]
    \centering
    \includegraphics[width=\linewidth]{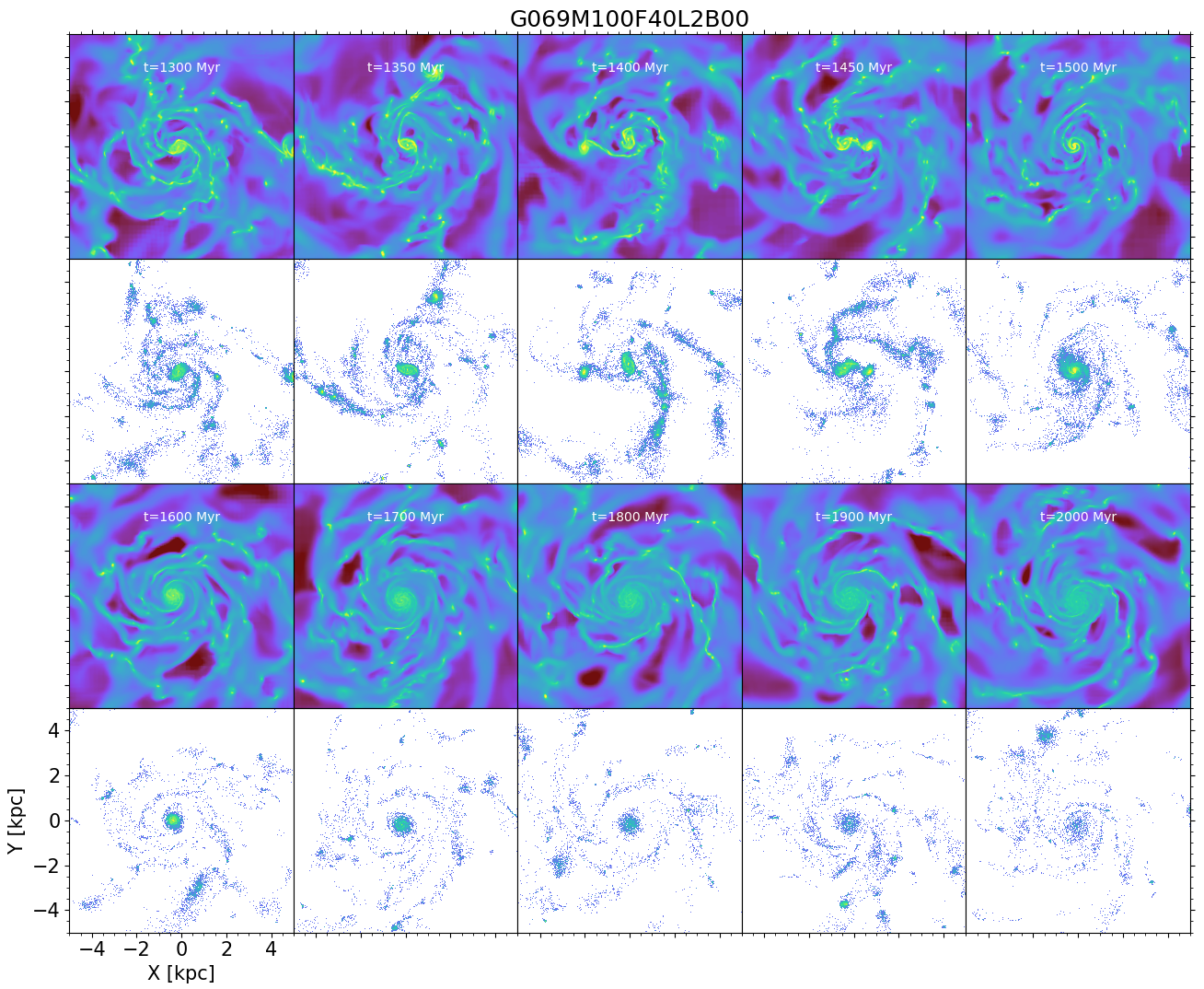}
    \caption{Density map of the newly formed stars of model G069 showing the interaction of the in situ formed stellar cluster and the bar region.}
    \label{fig:sf_map_G069}
\end{figure*}

The case of model G069 is much more relevant for this work because of its high gas fraction and the fact that it forms a bar after $\sim 315$ Myr. The interesting part is that the bar is destroyed after $\sim 1500$ Myr by a massive stellar cluster ($\sim 10^{7}$ \Msun) formed in situ during the run. In the bottom right panels of Fig.~\ref{fig:gas_map_nobar}, we observe that the bar has already formed after 500 Myr, with a higher gas density in the inner bar region, resulting in a high SFR. At t=1000 Myr, we see the formation of a central gas concentration, which marks the beginning of the building of the gas reservoir. Between t=1000 and 2000 Myr, a massive stellar cluster is formed in the outer part of the disc and is transported by dynamical friction to the centre, where it interacts with the bar and destroys it completely, as shown in Fig.~\ref{fig:sf_map_G069}. After the destruction of the bar, the SFR steadily decreases until the end of the simulation. Even though we did not classify model G069 as a barred model, its properties are very similar to the properties of barred models with higher stellar mass  until the moment of the interaction with the massive stellar cluster.
\end{appendix}
\end{document}